\pdfoutput=1

\documentclass[12pt,a4paper]{article}

\usepackage{afterpage} 
\usepackage{multirow}

\usepackage{ifthen} 
\usepackage{amsmath} 
\usepackage{bm} 
\newboolean{pdflatex}
\setboolean{pdflatex}{true} 

\newboolean{articletitles}
\setboolean{articletitles}{true} 

\newboolean{uprightparticles}
\setboolean{uprightparticles}{false} 

\newboolean{inbibliography}
\setboolean{inbibliography}{false} 

\usepackage{microtype}
\usepackage{lineno}  
\usepackage{xspace} 
\usepackage{caption} 

\usepackage{graphicx}  
\usepackage{color}
\usepackage{colortbl}
\graphicspath{{./figs/}} 

\usepackage{amsmath} 
\usepackage{amssymb}
\usepackage{amsfonts}
\usepackage{upgreek} 

\newcommand*\patchAmsMathEnvironmentForLineno[1]{%
\expandafter\let\csname old#1\expandafter\endcsname\csname #1\endcsname
\expandafter\let\csname oldend#1\expandafter\endcsname\csname
end#1\endcsname
 \renewenvironment{#1}%
   {\linenomath\csname old#1\endcsname}%
   {\csname oldend#1\endcsname\endlinenomath}%
}
\newcommand*\patchBothAmsMathEnvironmentsForLineno[1]{%
  \patchAmsMathEnvironmentForLineno{#1}%
  \patchAmsMathEnvironmentForLineno{#1*}%
}
\AtBeginDocument{%
\patchBothAmsMathEnvironmentsForLineno{equation}%
\patchBothAmsMathEnvironmentsForLineno{align}%
\patchBothAmsMathEnvironmentsForLineno{flalign}%
\patchBothAmsMathEnvironmentsForLineno{alignat}%
\patchBothAmsMathEnvironmentsForLineno{gather}%
\patchBothAmsMathEnvironmentsForLineno{multline}%
\patchBothAmsMathEnvironmentsForLineno{eqnarray}%
}

\usepackage{hyperref}    
\usepackage[all]{hypcap} 

\def\lhcb {\mbox{LHCb}\xspace}

\def\babar  {\mbox{BaBar}\xspace}
\def\belle  {\mbox{Belle}\xspace}
\def\cleo   {\mbox{CLEO}\xspace}

\def\MagUp {\mbox{\em Mag\kern -0.05em Up}\xspace}

\ifthenelse{\boolean{uprightparticles}}%
{

 \def\Ppi         {\ensuremath{\uppi}\xspace}

 \def\PDelta      {\ensuremath{\Delta}\xspace}                 
 \def\PXi      {\ensuremath{\Xi}\xspace}                 
 \def\PLambda      {\ensuremath{\Lambda}\xspace}                 
 \def\PSigma      {\ensuremath{\Sigma}\xspace}                 
 \def\POmega      {\ensuremath{\Omega}\xspace}                 
 \def\PUpsilon      {\ensuremath{\Upsilon}\xspace}                 
 
 \def\PB      {\ensuremath{\mathrm{B}}\xspace}                 
                  
 \def\PD      {\ensuremath{\mathrm{D}}\xspace}

 \def\PK      {\ensuremath{\mathrm{K}}\xspace}

 \def\Pb      {\ensuremath{\mathrm{b}}\xspace}                 
 \def\Pc      {\ensuremath{\mathrm{c}}\xspace}

 \def\Pi      {\ensuremath{\mathrm{i}}\xspace}

 \def\Ps      {\ensuremath{\mathrm{s}}\xspace}

}
{

 \def\Ppi         {\ensuremath{\pi}\xspace}

 \mathchardef\PDelta="7101
 \mathchardef\PXi="7104
 \mathchardef\PLambda="7103
 \mathchardef\PSigma="7106
 \mathchardef\POmega="710A
 \mathchardef\PUpsilon="7107
                  
 \def\PB      {\ensuremath{B}\xspace}                 
                  
 \def\PD      {\ensuremath{D}\xspace}

 \def\PK      {\ensuremath{K}\xspace}

 \def\Pb      {\ensuremath{b}\xspace}                 
 \def\Pc      {\ensuremath{c}\xspace}

 \def\Pi      {\ensuremath{i}\xspace}

 \def\Ps      {\ensuremath{s}\xspace}

}

\makeatletter
\ifcase \@ptsize \relax
  \newcommand{\miniscule}{\@setfontsize\miniscule{4}{5}}
\or
  \newcommand{\miniscule}{\@setfontsize\miniscule{5}{6}}
\or
  \newcommand{\miniscule}{\@setfontsize\miniscule{5}{6}}
\fi
\makeatother

\DeclareRobustCommand{\optbar}[1]{\shortstack{{\miniscule (\rule[.5ex]{1.25em}{.18mm})}
  \\ [-.7ex] $#1$}}

\def\squark    {{\ensuremath{\Ps}}\xspace}

\def\cquark    {{\ensuremath{\Pc}}\xspace}

\def\bquark    {{\ensuremath{\Pb}}\xspace}

\def\pion   {{\ensuremath{\Ppi}}\xspace}

\def\kaon    {{\ensuremath{\PK}}\xspace}
  \def\Kbar    {{\kern 0.2em\overline{\kern -0.2em \PK}{}}\xspace}

\def\KorKbar    {\kern 0.18em\optbar{\kern -0.18em K}{}\xspace}

\def\Kstarz  {{\ensuremath{\kaon^{*0}}}\xspace}
\def\Kstarzb {{\ensuremath{\Kbar{}^{*0}}}\xspace}

  \def\Dbar    {{\kern 0.2em\overline{\kern -0.2em \PD}{}}\xspace}

\def\DorDbar    {\kern 0.18em\optbar{\kern -0.18em D}{}\xspace}

\def\B       {{\ensuremath{\PB}}\xspace}
\def\Bbar    {{\ensuremath{\kern 0.18em\overline{\kern -0.18em \PB}{}}}\xspace}

\def\BorBbar    {\kern 0.18em\optbar{\kern -0.18em B}{}\xspace}
\def\BsorBsbar    {\kern 0.18em\optbar{\kern -0.18em B_s^0}{}\xspace}

\def\Bd      {{\ensuremath{\B^0}}\xspace}
\def\Bs      {{\ensuremath{\B^0_\squark}}\xspace}
\def\Bsb     {{\ensuremath{\Bbar{}^0_\squark}}\xspace}

  \def\Y#1S{\ensuremath{\PUpsilon{(#1S)}}\xspace}

\def\Lbar        {{\ensuremath{\kern 0.1em\overline{\kern -0.1em\PLambda}}}\xspace}
\def\LorLbar    {\kern 0.18em\optbar{\kern -0.18em \PLambda}{}\xspace}

\def\BF         {{\ensuremath{\cal B}}\xspace}

\def\BR         {\BF}

\def\to                 {\ensuremath{\rightarrow}\xspace}

\def\CP                {{\ensuremath{C\!P}}\xspace}
\def\CPT               {{\ensuremath{C\!PT}}\xspace}

\def\AT#1     {\ensuremath{A_{\mathrm{T}}^{#1}}\xspace}           

\def\C#1      {\ensuremath{\mathcal{C}_{#1}}\xspace}                       
\def\Cp#1     {\ensuremath{\mathcal{C}_{#1}^{'}}\xspace}                    
\def\Ceff#1   {\ensuremath{\mathcal{C}_{#1}^{\mathrm{(eff)}}}\xspace}        
\def\Cpeff#1  {\ensuremath{\mathcal{C}_{#1}^{'\mathrm{(eff)}}}\xspace}       
\def\Ope#1    {\ensuremath{\mathcal{O}_{#1}}\xspace}                       
\def\Opep#1   {\ensuremath{\mathcal{O}_{#1}^{'}}\xspace}                    

\newcommand{\ket}[1]{\ensuremath{|#1\rangle}}              

\newcommand{\tev}{\ifthenelse{\boolean{inbibliography}}{\ensuremath{~T\kern -0.05em eV}\xspace}{\ensuremath{\mathrm{\,Te\kern -0.1em V}}}\xspace}
\newcommand{\gev}{\ensuremath{\mathrm{\,Ge\kern -0.1em V}}\xspace}
\newcommand{\mev}{\ensuremath{\mathrm{\,Me\kern -0.1em V}}\xspace}
\newcommand{\kev}{\ensuremath{\mathrm{\,ke\kern -0.1em V}}\xspace}
\newcommand{\ev}{\ensuremath{\mathrm{\,e\kern -0.1em V}}\xspace}
\newcommand{\gevc}{\ensuremath{{\mathrm{\,Ge\kern -0.1em V\!/}c}}\xspace}
\newcommand{\mevc}{\ensuremath{{\mathrm{\,Me\kern -0.1em V\!/}c}}\xspace}
\newcommand{\gevcc}{\ensuremath{{\mathrm{\,Ge\kern -0.1em V\!/}c^2}}\xspace}
\newcommand{\gevgevcccc}{\ensuremath{{\mathrm{\,Ge\kern -0.1em V^2\!/}c^4}}\xspace}
\newcommand{\mevcc}{\ensuremath{{\mathrm{\,Me\kern -0.1em V\!/}c^2}}\xspace}

\def\mum  {\ensuremath{{\,\upmu\rm m}}\xspace}

\def\invpb {\ensuremath{\mbox{\,pb}^{-1}}\xspace}

\def\invfb   {\ensuremath{\mbox{\,fb}^{-1}}\xspace}

\newcommand{\chisqip}{\ensuremath{\chi^2_{\rm IP}}\xspace}

\def\gsim{{~\raise.15em\hbox{$>$}\kern-.85em
          \lower.35em\hbox{$\sim$}~}\xspace}
\def\lsim{{~\raise.15em\hbox{$<$}\kern-.85em
          \lower.35em\hbox{$\sim$}~}\xspace}

\newcommand{\Real}{\ensuremath{\mathcal{R}e}\xspace}
\newcommand{\Imag}{\ensuremath{\mathcal{I}m}\xspace}

\def\ptot       {\mbox{$p$}\xspace}
\def\pt         {\mbox{$p_{\rm T}$}\xspace}

\def\evtgen     {\mbox{\textsc{EvtGen}}\xspace}

\def\geant      {\mbox{\textsc{Geant4}}\xspace}

\def\photos     {\mbox{\textsc{Photos}}\xspace}

\def\pythia     {\mbox{\textsc{Pythia}}\xspace}

\def\tell1  {TELL1\xspace}
\def\ukl1   {UKL1\xspace}

\usepackage{cite} 
\usepackage{mciteplus}

\usepackage{longtable} 

\def\Kstars  {{\ensuremath{\kaon^{*}_0}}\xspace}

\newcommand{\dif}{\ensuremath{{\rm d}}\xspace}

\newcommand{\T}{\ensuremath{T}\xspace}

\newcommand{\Lambdab}{\ensuremath{\Lambda^0_b}\xspace}

\newcommand{\BsKstKst}{\ensuremath{\Bs\to\Kstarz\Kstarzb}\xspace}
\newcommand{\BdKstKst}{\ensuremath{\Bd\to\Kstarz\Kstarzb}\xspace}
\newcommand{\BsKpiKpi}{\ensuremath{\Bs\to K^+\pi^-K^-\pi^+}\xspace}
\newcommand{\BdPhiKst}{\ensuremath{\Bd\to\phi\Kstarz}\xspace}
\newcommand{\BsPhiKst}{\ensuremath{\Bs\to\phi\Kstarzb}\xspace}
\newcommand{\BdRhoKst}{\ensuremath{\Bd\to\rho\Kstarzb}\xspace}

\newcommand{\Lambdabppikst}{\ensuremath{\Lambdab\to p\pi^- K^- \pi^+}\xspace}

\newcommand{\BRof}[1]{\ensuremath{{\cal B}(#1)}\xspace}

\newcommand{\kkappa}{\ensuremath{K^*_0(800)}\xspace}
\newcommand{\khigh}{\ensuremath{K^*_0(1430)}\xspace}

\newcommand{\DLL}{\ensuremath{{\Delta \ln \mathcal{L}}}\xspace}

\newcommand{\swave}{{S--wave}\xspace}
\newcommand{\pwave}{{P--wave}\xspace}

\newcommand{\figref}[1]{Fig.~\ref{#1}}
\newcommand{\tabref}[1]{Table~\ref{#1}}

\newcommand{\secref}[1]{Sect.~\ref{#1}}
\renewcommand{\eqref}[1]{Eq.~(\ref{#1})}

\begin{document}

\renewcommand{\thefootnote}{\fnsymbol{footnote}}
\setcounter{footnote}{1}


\begin{titlepage}
\pagenumbering{roman}

\vspace*{-1.5cm}
\centerline{\large EUROPEAN ORGANIZATION FOR NUCLEAR RESEARCH (CERN)}
\vspace*{1.5cm}
\hspace*{-0.5cm}
\begin{tabular*}{\linewidth}{lc@{\extracolsep{\fill}}r}
\ifthenelse{\boolean{pdflatex}}
{\vspace*{-2.7cm}\mbox{\!\!\!\includegraphics[width=.14\textwidth]{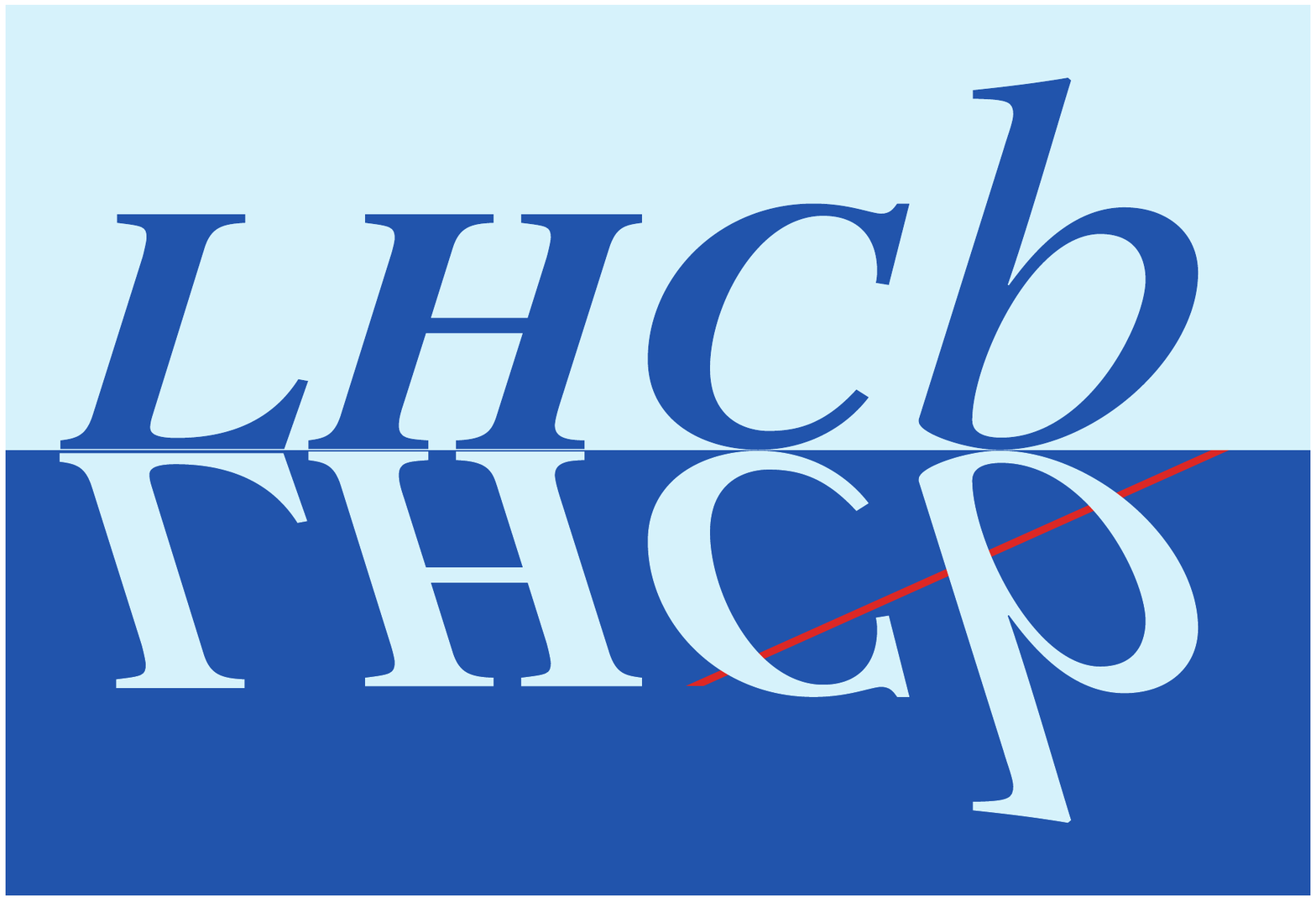}} & &}%
{\vspace*{-1.2cm}\mbox{\!\!\!\includegraphics[width=.12\textwidth]{lhcb-logo.eps}} & &}%
\\
 & & CERN-PH-EP-2015-058 \\  
 & & LHCb-PAPER-2014-068 \\  
 & & June 18, 2015 \\ 
 & & \\
\end{tabular*}

\vspace*{3.0cm}

{\bf\boldmath\huge
\begin{center}
  Measurement of \CP asymmetries and polarisation fractions in $\Bs\to\Kstarz\Kstarzb$ decays 
\end{center}
}

\vspace*{1.0cm}

\begin{center}
The LHCb collaboration\footnote{Authors are listed at the end of this paper.}
\end{center}

\vspace{\fill}

\begin{abstract}
  \noindent
  An angular analysis of the decay \BsKstKst
  is performed using $pp$ collisions corresponding to an integrated luminosity of $1.0\,{\rm fb}^{-1}$ 
  collected by the LHCb experiment at a centre-of-mass energy $\sqrt{s} = 7\,{\rm TeV}$. A combined angular and mass analysis 
  separates six helicity amplitudes and allows the measurement of
  the longitudinal polarisation fraction $f_L = 0.201 \pm 0.057 {\ \rm(stat.)} \pm 0.040{\ \rm(syst.)}$ for the $\Bs\to K^*(892)^0 \Kbar{}^*(892)^0$ decay.
  A large scalar contribution from the $K^{*}_{0}(1430)$ and $K^{*}_{0}(800)$ resonances is found, allowing the determination of additional \CP asymmetries.
  Triple product and direct $C\!P$ asymmetries are determined to be compatible with the Standard Model expectations. The branching fraction 
  $\BR(\Bs\to K^*(892)^0 \Kbar{}^*(892)^0)$ is measured to be $(10.8 \pm 2.1 {\ \rm (stat.)} \pm 1.4 {\ \rm (syst.)} \pm 0.6 \ (f_d/f_s) ) \times 10^{-6}$.
\end{abstract}

\vspace*{2.0cm}

\begin{center}
  Submitted to JHEP 
\end{center}

\vspace{\fill}

{\footnotesize 
\centerline{\copyright~CERN on behalf of the \lhcb collaboration, license \href{http://creativecommons.org/licenses/by/4.0/}{CC-BY-4.0}.}}
\vspace*{2mm}

\end{titlepage}


\newpage
\setcounter{page}{2}
\mbox{~}
%
%
%
%

\cleardoublepage


\renewcommand{\thefootnote}{\arabic{footnote}}
\setcounter{footnote}{0}



\pagestyle{plain} 
\setcounter{page}{1}
\pagenumbering{arabic}


%

\section{Introduction}
\label{sec:Introduction}

The \BsKstKst decay is mediated by a $b\to s d \bar{d}$ flavour-changing neutral current
(FCNC) transition, which in the Standard Model (SM) 
proceeds through loop diagrams at leading order. This decay has been discussed in the literature as a possible field for precision
tests of the SM predictions, when it is considered in association with its U-spin symmetric
channel \BdKstKst~\cite{fleischer99,ciucini,matias}. In the SM, the expected \CP violation
in the former is very small, $\mathcal{O}(\lambda^2)$, with approximate cancellation between 
the mixing and the decay CKM phases \cite{london_new}. When a scalar meson background is allowed,
in addition to the vector-vector meson states, six independent helicities contribute \cite{london_new}.

In this paper, a search for non-SM 
electroweak amplitudes is reported in the decay $B^0_s \rightarrow K^+\pi^-K^-\pi^+$,
with $K\pi$ mass close to the $K^{*}(892)^0$ mass,
through the measurement of all \CP-violating observables accessible when
the flavour of the bottom-strange meson is not identified.
These observables include triple products (TPs) and other \CP-odd quantities \cite{TP_londonDatta},
many of which are, as yet, experimentally unconstrained. 
Triple products are \T-odd observables having the generic structure 
$\boldsymbol{v_1} \cdot (\boldsymbol{v_2} \times \boldsymbol{v_3})$ where
$\boldsymbol{v_i}$ is the spin or momentum of a final-state particle.
In vector-vector final states of \B mesons they take the form 
$\boldsymbol{q} \cdot (\boldsymbol{\epsilon_1} \times \boldsymbol{\epsilon_2})$ where $\boldsymbol{q}$ is the momentum of
one of the final vector mesons and $\boldsymbol{\epsilon_1}$ and $\boldsymbol{\epsilon_2}$ are their
respective polarisations. Triple products are also meaningful when one of the
final particles is a scalar meson. 

Theoretical predictions based on perturbative QCD for the decay of \B mesons into scalar-vector final states $K^*_0(1430) \Kbar{}^*(892)^0$ have been recently investigated, yielding branching fractions comparable to those of vector-vector final states~\cite{Liu}, which have been previously available \cite{Beneke}. 
The \BsKstKst decay was first observed with 
35\invpb of LHCb data~\cite{KstKstPaper} reporting
the measurement of the branching fraction and an angular analysis.
A remarkably low longitudinal polarisation fraction was observed, 
compatible with that found for the similar decay $B^0_s \rightarrow \phi\phi$~\cite{phiphi_paper}, and
at variance with that observed in the mirror channel \BdKstKst~\cite{Babar} and with some predictions from QCD factorisation~\cite{Beneke,matias_2013}.

An updated analysis of the $B^0_s \rightarrow K^+\pi^-K^-\pi^+$ final state is reported
in this publication, in the mass
window of $\pm150$\mevcc around the $K^{*}(892)^0$ (hereafter referred to as \Kstarz) mass for $K^+\pi^-$ and
$K^-\pi^+$ pairs. A description of the \CP observables is provided in
Section \ref{sec:strategy}, the LHCb apparatus is summarised in Section \ref{sec:detector}, and the
data sample described in Section \ref{sec:selection}.
Triple products and direct \CP asymmetries are determined in Section \ref{sec:tpa}. 
A measurement of the various amplitudes contributing to $\Bs \rightarrow K^+\pi^-K^-\pi^+$ 
is performed in Section \ref{sec:angular}, under the assumption of \CP conservation. These include the polarisation fractions for the 
vector-vector mode \BsKstKst. In light of these results, the measurement of the branching fraction
\BRof\BsKstKst is updated in Section \ref{sec:br}. Conclusions are summarised in Section \ref{sec:conclusions}.
These studies are performed using 1.0\invfb of $pp$ collision data from the LHC at a centre-of-mass energy of $\sqrt{s}=7$\tev and recorded with the LHCb detector.

\section{Analysis strategy}
\label{sec:strategy}

Considering only the \swave ($J_{1,2}=0$) and \pwave ($J_{1,2}=1$) production of the $K\pi$ pairs, 
with $J _{1,2}$ the angular momentum of the respective $K\pi$ combination, 
the decay $\Bs \rightarrow (K^+\pi^-)_{J_1} (K^-\pi^+)_{J_2}$ can be described in 
terms of six helicity decay amplitudes.
A two-dimensional fit to the $K^+\pi^-$ and $K^-\pi^+$ mass spectra, for masses up to
the $\kaon^{*}_J(1430)^{0}$ resonances, finds a small contribution ($<1\%$) of tensor
amplitudes when projected onto the $K\pi$ mass interval used in this analysis, and thus
these amplitudes are not considered.
Three of the above amplitudes describe the decay into two vector mesons, commonly referred to as \pwave amplitudes, 
$\Bs\to V_1 V_2$ with $V_1=\Kstarz$ and $V_2=\Kstarzb$, with the 
subsequent two-body strong-interaction decay of each of the vector mesons into a $K\pi$ pair. Each amplitude corresponds
to a different helicity ($L_z = 0, +1, -1$) of the vector mesons in the final state with respect to their 
relative momentum direction, $H_0$, $H_{+}$ and $H_{-}$.
It is useful to write the decay rate in terms of the amplitudes in the transversity basis,
\begin{equation}
A_0 = H_0,\ \ \ \ \   A_{\parallel} = \frac{1}{\sqrt{2}} \left( H_+ + H_- \right)\ \ \ \ \  {\rm and} \ \ \ \ \ A_{\perp} = \frac{1}{\sqrt{2}} \left( H_+ - H_- \right),
\end{equation}
since, unlike the helicity amplitudes, they correspond to states with definite \CP eigenvalues ($\eta_{\parallel} = \eta_0 = 1$ and $\eta_{\perp} = -1$).
The \pwave amplitudes are assumed to have a relativistic Breit-Wigner dependence on the $K\pi$ invariant mass.

In addition, contributions arising from decays into scalar resonances or non-resonant $K\pi$ pairs 
need to be taken into account within the mass window indicated above. The amplitudes describing this \swave
configuration are $A_{V\!S}$, $A_{SV}$ and $A_{SS}$, corresponding to the following decays\footnote{Note that both \Bs and \Bsb can decay into these final states.}
\begin{equation}
\renewcommand{\arraystretch}{1.5}
\setlength{\arraycolsep}{1pt}
  \begin{array}{cccl}
    {A}_{V\!S} &\ \ \ :\ \ \ &  \Bs &\rightarrow \Kstarz  {(K^- \pi^+)}_0,  \\
    {A}_{SV} &\ \ \ :\ \ \ & \Bs &\rightarrow {(K^+ \pi^-)}_0  \Kstarzb \quad{\rm and}  \\
    A_{SS} &\ \ \ :\ \ \ & \Bs &\rightarrow {(K^+\pi^-)}_0 {(K^- \pi^+)}_0, 
  \end{array}
  \label{eq:SV_amps}
\end{equation}
where the subscript denotes the relative orbital angular momentum, $J$, 
of the pair. 
The scalar combinations $(K\pi)_0$ are described by a superposition 
of a broad low-mass structure related to the $K^*_0(800)$ resonance~\cite{DescotesGenon} and a component describing the 
$K^*_0(1430)$ resonance.

Unlike the $\Kstarz\Kstarzb$ final state, the \swave configurations $SV$ and $V\!S$ defined in \eqref{eq:SV_amps}
do not correspond to \CP eigenstates. However, one may consider the following superpositions
\begin{eqnarray}
  \ket{s^+} &=& \frac{1}{\sqrt{2}} \left( \ket{\Kstarz (K^- \pi^+)_0} + \ket{(K^+ \pi^-)_0 \Kstarzb} \right) \quad{\rm and} \nonumber \\
  \ket{s^-} &=& \frac{1}{\sqrt{2}} \left( \ket{\Kstarz (K^- \pi^+)_0} - \ket{(K^+ \pi^-)_0 \Kstarzb} \right),
\end{eqnarray}
\noindent which are indeed \CP eigenstates with opposite \CP parities ($\eta_{s^+}=-1$ and $\eta_{s^-}=+1$). 
Therefore, it is possible to write the full decay amplitude in terms of \CP-odd and \CP-even amplitudes
(the $S\!S$ final configuration is a \CP eigenstate with $\eta_{SS}=1$) by defining
\begin{equation}
  A_s^+ = \frac{1}{\sqrt{2}}\left(A_{V\!S} + A_{SV} \right) \quad{\rm and} \quad A_s^- = \frac{1}{\sqrt{2}}\left(A_{V\!S} - A_{SV} \right).
\label{eq:defaspasm}
\end{equation}

\subsection{Angular distribution}

\begin{figure}
  \centering
  \includegraphics[width=0.5\textwidth]{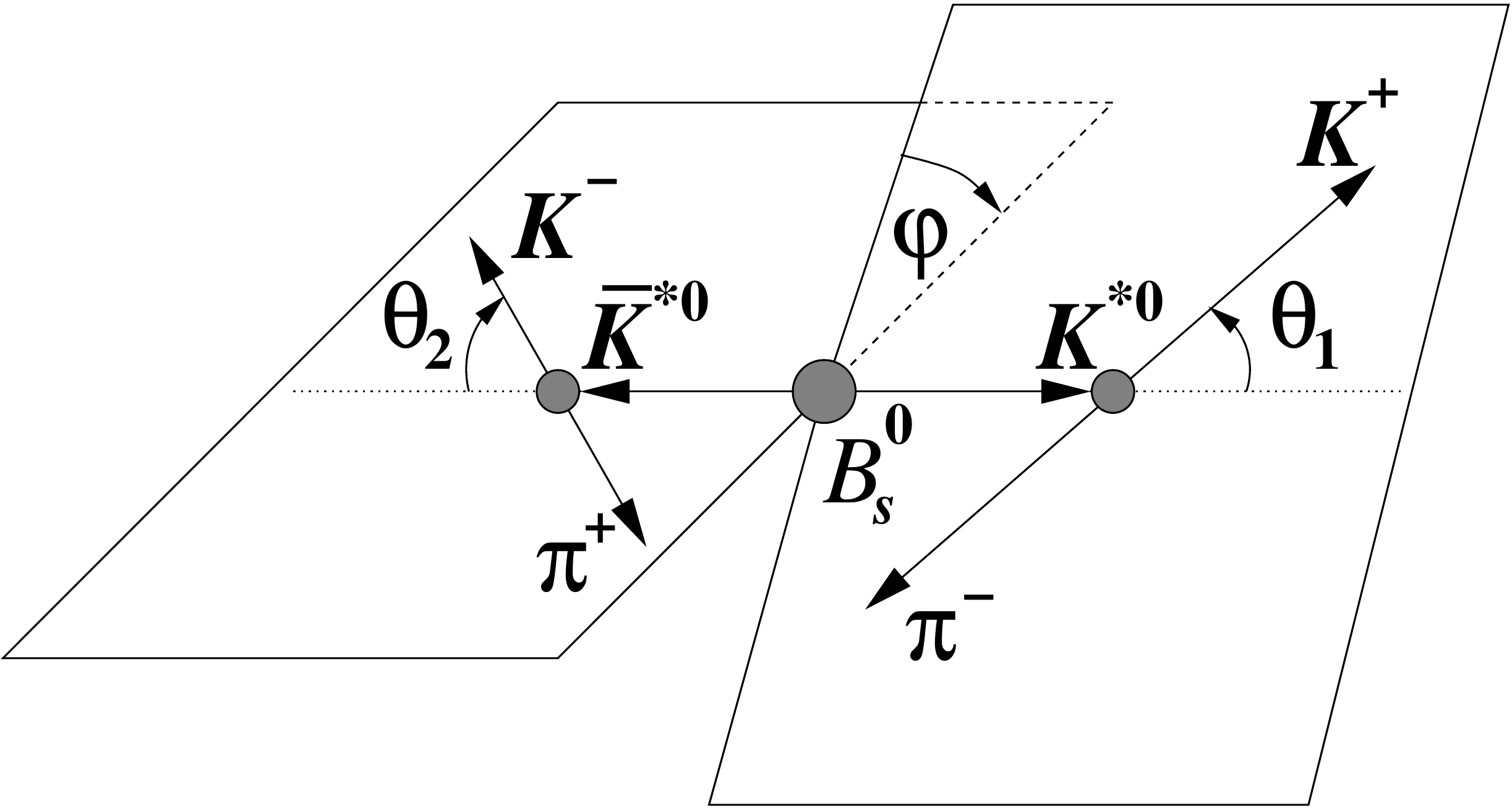}
  \caption{ Definition of the angles involved in the analysis of \BsKstKst decays.
 }
\label{fig:anglesdef}
\end{figure}

The angles describing the decay, $\Omega\equiv\{\theta_1,\theta_2,\varphi\}$, 
are shown in \figref{fig:anglesdef}, where $\theta_{1(2)}$ is the 
angle between the direction of $K^{+(-)}$ meson and the direction opposite to the $B$-meson momentum 
in the rest frame of $V_{1(2)}$ and $\varphi$ is the angle between the decay planes of the two vector mesons in the 
$B$-meson rest frame. 
In this angular basis, the differential decay rate describing this process is expressed as~\cite{angular_distribution}
\begin{eqnarray}
 \frac{\dif^6\Gamma}{ \dif \Omega \, \dif m_1\, \dif m_2\, \dif t} &=& N\ \Big|\ \Big(\mathcal{A}_0(t) \cos\theta_1 \cos\theta_2  +  \frac{{\mathcal{A}}_{\parallel}(t)}{\sqrt {2}} \sin\theta_1 \sin\theta_2 \cos\varphi \nonumber \\
   && \ \ \ \ \ \ \ \  + i \  \frac{{\mathcal{A}}_{\perp}(t)}{\sqrt {2}} \sin\theta_1 \sin\theta_2 \sin\varphi \ \Big) \mathcal{M}_1(m_1) \mathcal{M}_1(m_2) \nonumber \\
   && \ \ \ \ - \frac{\mathcal{A}_{s}^+(t)}{\sqrt{6}} \Big(\cos\theta_1 \mathcal{M}_1(m_1) \mathcal{M}_0(m_2) - \cos\theta_2 \mathcal{M}_0(m_1) \mathcal{M}_1(m_2)\Big) \nonumber \\
   && \ \ \ \ - \frac{\mathcal{A}_{s}^-(t)}{\sqrt{6}} \Big(\cos\theta_1 \mathcal{M}_1(m_1) \mathcal{M}_0(m_2) + \cos\theta_2 \mathcal{M}_0(m_1) \mathcal{M}_1(m_2)\Big) \nonumber \\
   && \ \ \ \ - \frac{\mathcal{A}_{ss}}{3} \mathcal{M}_0(m_1) \mathcal{M}_0(m_2)    \ \ \Big|^2 ,
\label{eq:rate_timedep}
\end{eqnarray}
\noindent where the different dependences of \pwave and \swave amplitudes on the two-body masses 
$m_1 \equiv M(K^+ \pi^-)$ and $m_2 \equiv M(K^- \pi^+)$  
have been made explicit in terms of the mass propagators $\mathcal{M}_{1,0}(m)$ and
$N$ is an overall normalisation constant.
The time evolution induced 
by \Bs--\Bsb mixing is encoded in the time-dependence of the amplitudes $\mathcal{A}_k(t)$ ($k=0, \parallel, \perp, s^+, s^-, ss$)
\begin{equation}
	\mathcal{A}_k(t) = g_+(t) A_k + \left(\frac{q}{p}\right) g_{-}(t) \bar{A}_k , 
	\label{eq:amplitude_timedep}
\end{equation}
where $A_k \equiv \mathcal{A}_k(t=0)$ and the time-dependent functions $g_{\pm}(t)$ are given by
\begin{equation}
	g_{\pm}(t) = \frac{1}{2} \left(  e^{-i m_H t - \frac{1}{2} \Gamma_H t} \pm e^{-i m_L t - \frac{1}{2} \Gamma_L t } \right)
\end{equation}
with $\Gamma_{L,H}$ ($m_{L,H}$) being the width (mass) of the \Bs light ($L$) and heavy ($H$) mass eigenstates. 

The decay rate of the \CP-conjugated process, $\Bsb\to (K^- \pi^+) (K^+ \pi^-)$, can be obtained by 
exchanging each amplitude $A_k$ by $\eta_k \bar{A}_k$, where $\eta_k$ is the \CP eigenvalue
of the final state described by $A_k$~\cite{london_new}.
In this paper, due to the limited size of the available data sample, no attempt is made to identify the flavour of the initial \Bs meson at production, thus suppressing 
the sensitivity to direct and mixing-induced \CP asymmetries. Nevertheless, \CP violation can still be 
studied through the measurement of triple product asymmetries and \swave-induced direct \CP asymmetries.

\subsection{Triple product asymmetries}
\label{sec:tpa_def}
Two TPs can be defined in \B meson decays into pairs of vector particles~\cite{TP_londonDatta,gronau,london_new},
\begin{eqnarray}
	T_1 &=& (\hat{n}_{V_1} \times \hat{n}_{V_2})\cdot \hat{p}_{V_1} = \sin\varphi \quad{\rm and} \\
	\nonumber \\
	T_2 &=& 2(\hat{n}_{V_1} \cdot \hat{n}_{V_2})(\hat{n}_{V_1} \times \hat{n}_{V_2})\cdot \hat{p}_{V_1} = \sin2\varphi,	 
\end{eqnarray}
where $\hat{n}_{V_i}$ ($i=1,2$) is a unit vector perpendicular to the $V_i$ decay plane and $\hat{p}_{V_1}$ is a unit 
vector in the direction of $V_1$ in the \Bs rest frame. The observable asymmetries associated with these TPs can be 
calculated from integrations of the differential decay rate as~\cite{gronau}
\begin{eqnarray}
  a_T^1(t) &\equiv& \frac{\Gamma(\cos\theta_1 \cos\theta_2\sin\varphi>0, t) - \Gamma(\cos\theta_1 \cos\theta_2 \sin\varphi<0, t)}{\Gamma(\cos\theta_1 \cos\theta_2\sin\varphi>0, t) + \Gamma(\cos\theta_1 \cos\theta_2 \sin\varphi<0, t)} \quad{\rm and} \label{formula:TP_1} \\
  &&  \nonumber \\
  a_T^2(t) &\equiv& \frac{\Gamma(\sin2\varphi>0, t) - \Gamma(\sin2\varphi<0, t)}{\Gamma(\sin2\varphi>0, t) + \Gamma(\sin2\varphi<0, t)}.   \label{formula:TP_2}
\end{eqnarray}
Nonzero TP asymmetries appear either due to a \T-violating phase or a \T-conserving phase in conjunction with final-state interactions.

When these asymmetries are measured in a sample where the production flavour is not identified, they become ``true'' \CP-violating asymmetries,
assuming that \CPT is conserved. This is manifest when $a_T^1$ and $a_T^2$ are written in terms of the amplitudes defining the
decay rate in \eqref{eq:rate_timedep},
\begin{eqnarray}
  a_T^1 &=& \frac{2\sqrt{2}}{\pi} \frac{1}{\mathcal{D}} \Imag(\mathcal{A}_{\perp}\mathcal{A}_0^*) \quad{\rm and}  \\
  &&  \nonumber \\
  a_T^2 &=& -\frac{4}{\pi}\frac{1}{\mathcal{D}} \Imag(\mathcal{A}_{\perp}\mathcal{A}_{\parallel}^*)  ,
\end{eqnarray}
with $\mathcal{D} = |\mathcal{A}_0|^2+|\mathcal{A}_{\parallel}|^2+|\mathcal{A}_{\perp}|^2+|\mathcal{A}_s^{+}|^2+|\mathcal{A}_s^{-}|^2 +|\mathcal{A}_{ss}|^2$. Taking into account these expressions and 
that the decay rate contribution associated with the \CP-odd amplitude $\mathcal{A}_{\perp}$ changes sign
under the \CP transformation, the asymmetries measured in the untagged sample, ${A}_T^i$, are proportional 
to the \CP-violating interference terms $\Imag(\mathcal{A}_{\perp} \mathcal{A}_{0,\parallel}^* - \bar{\mathcal{A}}_{\perp} \bar{\mathcal{A}}_{0,\parallel}^*)$.
Using \eqref{eq:amplitude_timedep}, these terms can be written as
\begin{align}
	\Imag(\mathcal{A}_{\perp} \mathcal{A}_{0,\parallel}^* - \bar{\mathcal{A}}_{\perp} \bar{\mathcal{A}}_{0,\parallel}^*) = \frac{1}{2} e^{-\Gamma_s t} 
	\bigg[ &\, & \Imag&({A}_{\perp} {A}_{0,\parallel}^* - \bar{{A}}_{\perp} \bar{{A}}_{0,\parallel}^*) & \cosh&\left(\frac{\Delta\Gamma_s}{2} t\right)  \nonumber \\
	& + &\Imag&[(\bar{{A}}_{\perp} {A}_{0,\parallel}^* + {A}_{\perp}^* \bar{{A}}_{0,\parallel}) e^{-i\phi_{\rm mix}}] & \sinh&\left(\frac{\Delta\Gamma_s}{2} t\right) \bigg],
\label{eq:TP_timedep}
\end{align}
where $\Delta\Gamma_s \equiv \Gamma_L-\Gamma_H$, $\Gamma_s \equiv (\Gamma_L + \Gamma_H)/{2}$ and $\phi_{\rm mix}$ is the phase in \Bs--\Bsb mixing.
The coefficients $\Imag({A}_{\perp} {A}_{0,\parallel}^* - \bar{{A}}_{\perp} \bar{{A}}_{0,\parallel}^*)$ and $\Imag[(\bar{{A}}_{\perp} {A}_{0,\parallel}^* + {A}_{\perp}^* \bar{{A}}_{0,\parallel}) e^{-i\phi_{\rm mix}}]$ are TP and mixing-induced TP asymmetries, respectively, and are \CP-violating quantities~\cite{london_new}. In the analysis presented in this paper, only the time-integrated asymmetries 
\begin{eqnarray}
    A_T^1 &=& \frac{2\sqrt{2}}{\pi} \frac{1}{{D}} \int \Imag(\mathcal{A}_{\perp}\mathcal{A}_0^*) \, {\rm d}t \quad{\rm and}  \\
    &&  \nonumber \\
    A_T^2 &=& -\frac{4}{\pi}\frac{1}{{D}} \int \Imag(\mathcal{A}_{\perp}\mathcal{A}_{\parallel}^*) \, {\rm d}t ,
\end{eqnarray}
are measured ($D = \int \mathcal{D} \, {\rm d}t$), with no identification of initial \Bs flavour. Thus \CP-violating linear combinations 
of the above observables are accessible.

When the \swave contribution is taken into account, two additional \CP-even amplitudes, $\mathcal{A}_s^-$ and $\mathcal{A}_{ss}$, interfere with $\mathcal{A}_{\perp}$, and give rise to two additional \CP-violating terms.
Further asymmetric integrations of the decay rate, analogous to those in~\cite{gronau}, lead to the following observables
\begin{align}
{A}_T^{3} & \equiv \frac{\Gamma((\cos\theta_1 + \cos\theta_2)\sin \varphi > 0) - \Gamma((\cos\theta_1+ \cos\theta_2)\sin \varphi < 0)}{\Gamma((\cos\theta_1+ \cos\theta_2)\sin \varphi > 0) + \Gamma((\cos\theta_1+ \cos\theta_2)\sin \varphi < 0)} \nonumber \\
&= \frac{32}{5\pi \sqrt{3}}\frac{1}{\mathcal{D}} \int\Imag\left( \left( \mathcal{A}_{\perp}\mathcal{A}_s^{-*} -   \bar{\mathcal{A}}_{\perp} \bar{\mathcal{A}}_s^{-*} \right)  \mathcal{M}_1(m) \mathcal{M}_0^*(m)\right) \mathrm{d}m  
\label{formula:TP_3}
 \\
\intertext{and} 
{A}_T^{4} & \equiv \frac{\Gamma(\sin \varphi > 0) - \Gamma(\sin \varphi < 0)}{\Gamma(\sin \varphi > 0) + \Gamma(\sin \varphi < 0)} \nonumber \\
& = \frac{3\pi}{4\sqrt{2}}\frac{1}{\mathcal{D}} \int\Imag\left( \left( \mathcal{A}_{\perp}\mathcal{A}_{ss}^{*} - \bar{\mathcal{A}}_{\perp}\bar{\mathcal{A}}_{ss}^{*}  \right)\mathcal{M}_1(m)\mathcal{M}_0^*(m)\right)\mathrm{d}m  ,
\label{formula:TP_4}
\end{align}
where the mass integration extends over the chosen $K\pi$ mass window. It is performed 
over the product of mass propagators of different resonances, times specific \CP-violating observables involving $\mathcal{A}_{\perp}$.

Since $\mathcal{A}_s^+$ is also \CP-odd, its interference terms with the \CP-even amplitudes change
sign under \Bs to \Bsb interchange.
Consequently, four new \CP-violating asymmetries 
are accessible from $\Bs\rightarrow K^+\pi^-K^-\pi^+$ decays, 
\begin{align}
{A}_D^{1} & \equiv \frac{\Gamma(\cos\theta_1\cos\theta_2(\cos\theta_1-\cos\theta_2) > 0) - \Gamma(\cos\theta_1\cos\theta_2(\cos\theta_1-\cos\theta_2) < 0)}{\Gamma(\cos\theta_1\cos\theta_2(\cos\theta_1-\cos\theta_2) > 0) + \Gamma(\cos\theta_1\cos\theta_2(\cos\theta_1-\cos\theta_2) < 0)} \nonumber \\
& = \frac{\sqrt{2}}{5\sqrt{3}} \frac{1}{\mathcal{D}} \left[ 9\int \Real\left( (\mathcal{A}_{s}^+\mathcal{A}_0^* - \bar{\mathcal{A}}_s^+ \bar{\mathcal{A}}_0^*) \mathcal{M}_0(m)\mathcal{M}_1^*(m) \right) \mathrm{d} m \right. \nonumber \\
& \left. \ \ \ \ \ \ \ \ \ \ \ \ +5 \int \Real\left( (\mathcal{A}_{s}^+\mathcal{A}_{ss}^* - \bar{\mathcal{A}}_s^+ \bar{\mathcal{A}}_{ss}^*) \mathcal{M}_1(m)\mathcal{M}_0^*(m) \right) \mathrm{d} m \right]  , 
\label{formula:DCPA_1}
 \\
\nonumber \\
{A}_D^{2} & \equiv \frac{\Gamma((\cos\theta_1-\cos\theta_2)\cos\varphi > 0) - \Gamma((\cos\theta_1-\cos\theta_2)\cos\varphi < 0)}{\Gamma((\cos\theta_1-\cos\theta_2)\cos\varphi > 0) + \Gamma((\cos\theta_1-\cos\theta_2)\cos\varphi < 0)} \nonumber \\ 
&= - \frac{32}{5\pi \sqrt{3}}\frac{1}{\mathcal{D}} \int  \Real\left( (\mathcal{A}_s^+\mathcal{A}_{\parallel}^* - \bar{\mathcal{A}}_{s}^+\bar{\mathcal{A}}_{\parallel}^*) \mathcal{M}_0(m)\mathcal{M}_1^*(m) \right)  \mathrm{d} m  ,
\label{formula:DCPA_2}
\\ \nonumber \\
{A}_D^{3} & \equiv \frac{\Gamma((\cos\theta_1-\cos\theta_2)> 0) - \Gamma((\cos\theta_1-\cos\theta_2) < 0)}{\Gamma((\cos\theta_1-\cos\theta_2) > 0) + \Gamma((\cos\theta_1-\cos\theta_2) < 0)} \nonumber \\ 
& = \frac{2\sqrt{2}}{5\sqrt{3}} \frac{1}{\mathcal{D}} \left[ 3\int  \Real\left( (\mathcal{A}_{s}^+\mathcal{A}_0^* - \bar{\mathcal{A}}_s^+ \bar{\mathcal{A}}_0^*) \mathcal{M}_0(m)\mathcal{M}_1^*(m) \right) \mathrm{d} m \right. \nonumber \\
& \left. \ \ \ \ \ \ \ \ \ \ \ \ + 5 \int  \Real\left( (\mathcal{A}_{s}^+\mathcal{A}_{ss}^* - \bar{\mathcal{A}}_s^+ \bar{\mathcal{A}}_{ss}^*) \mathcal{M}_1(m)\mathcal{M}_0^*(m) \right) \mathrm{d} m \right] 
\label{formula:DCPA_3}
\\
\intertext{and}
{A}_D^{4} & \equiv \frac{\Gamma((\cos^2\theta_1-\cos^2\theta_2)> 0) - \Gamma((\cos^2\theta_1-\cos^2\theta_2) < 0)}{\Gamma((\cos^2\theta_1-\cos^2\theta_2) > 0) + \Gamma((\cos^2\theta_1-\cos^2\theta_2) < 0)} \nonumber \\ 
& = \frac{1}{\mathcal{D}}\Real\left(\mathcal{A}_{s}^+\mathcal{A}_{s}^{-*} - \bar{\mathcal{A}}_{s}^+\bar{\mathcal{A}}_{s}^{-*}  \right)  .
\label{formula:DCPA_4}
\end{align}
Some of these terms have the form $\Real(\mathcal{A}_{s}^+ \mathcal{A}_k^* - \bar{\mathcal{A}}_s^+ \bar{\mathcal{A}}_k^*)$, with $k=0,\parallel,s^-,ss$,
which is characteristic of direct \CP asymmetries. 

As shown above, TP and several direct \CP asymmetries are accessible from untagged 
\BsKstKst decays, provided that a scalar $K\pi$ background component is present.
These \CP-violating observables are sensitive to the contributions of 
FCNC processes induced by neutral scalars, which are present, for example, in models with an extended Higgs sector. 
Constraints on possible FCNC couplings of Higgs scalars have been recently examined~\cite{extHiggs1,extHiggs2}. Non-zero values of $A_T^{i}$ or $A_D^{i}$ 
would allow the characterisation of those operators contributing to the effective Hamiltonian. In particular an enhanced 
contribution from $A_D^{1,2,3}$ with respect to the other observables, would reveal stronger $(V-A)\times(V+A)$ ($LR$) and 
$(V+A)\times(V-A)$ ($RL$) components with respect to $RR$ and $LL$ operators in the above models~\cite{london_new}.

\subsection{Angular analysis}

Assuming that no \CP violation arises in this decay, an angular analysis of the decay products
determines the polarisation fractions of the \BsKstKst decay and the contribution of the 
various \swave amplitudes. The time-integrated decay rate can be expressed as
\begin{equation}
 \frac{\dif^5\Gamma}{ \dif \Omega\, \dif m_1\, \dif m_2} = \hat{N}  \sum_{n=1}^{21} K_n(m_1,m_2) F_n(\Omega)  ,
\label{eq:rate_timeint}
\end{equation}
where the functions $K_n$ contain the dependence on the amplitudes entering the decay, 
with their corresponding mass propagators, and $\hat{N}$ is an overall normalisation constant. 
The $K_n$ functions are given in~\tabref{tab:funcs_timeint} together with
the decay angle functions $F_n$. 
All terms proportional to the TP and \swave-induced \CP asymmetries
($n=5,6,21$ and the symmetric $A_s^+ \leftrightarrow A_s^-$ terms in $n=8-11,13-16$) 
cancel under the assumption of \CP conservation.

\begin{table}[p]
  \small
  \caption{Untagged time-integrated terms used in the analysis, under the assumption of no \CP violation.}
  \label{tab:funcs_timeint}
 \renewcommand{\arraystretch}{1.5}
  \centering
  \begin{tabular}{ccc}
    \hline
    $n$ & $K_n$ & $F_n$ \\
    \hline
    1 & $\frac{1}{\Gamma_L}{|A_0|}^2 |\mathcal{M}_1(m_1)|^2 |\mathcal{M}_1(m_2)|^2$ & $\cos^2\theta_1 \cos^2\theta_2$ \\
    2 & $\frac{1}{\Gamma_L}{|A_{\parallel}|}^2 |\mathcal{M}_1(m_1)|^2 |\mathcal{M}_1(m_2)|^2$ & $\frac{1}{2}\sin^2\theta_1 \sin^2\theta_2 \cos^2\varphi$ \\
    3 & $\frac{1}{\Gamma_H}{|A_{\perp}|}^2 |\mathcal{M}_1(m_1)|^2 |\mathcal{M}_1(m_2)|^2$ & $\frac{1}{2}\sin^2\theta_1 \sin^2\theta_2 \sin^2\varphi$ \\
    4 & $\frac{1}{\Gamma_L}{|A_{\parallel}||A_{0}|} \cos{\delta_{\parallel}}  |\mathcal{M}_1(m_1)|^2 |\mathcal{M}_1(m_2)|^2$ & $\frac{1}{2\sqrt{2}}\sin 2\theta_1 \sin 2\theta_2 \cos\varphi$ \\
    5 & $0$ & $-\frac{1}{2\sqrt{2}}\sin 2\theta_1 \sin 2\theta_2 \sin\varphi$ \\
    6 & $0$ & $-\frac{1}{2} \sin^2\theta_1 \sin^2\theta_2 \sin 2\varphi$ \\
    \hline
    7 & $\frac{1}{2}  (\frac{{|A_s^+|}^2}{\Gamma_H} + \frac{{|A_s^-|}^2}{\Gamma_L} ) |\mathcal{M}_1(m_1)|^2 |\mathcal{M}_0(m_2)|^2$ & $\frac{1}{3} \cos^2\theta_1$ \\
    8 & $\frac{1}{\sqrt{2}}  \frac{1}{\Gamma_L} {|A_s^-|} {|A_{0}|} \Real( e^{i{\delta_s^-}} \mathcal{M}_1^*(m_2) \mathcal{M}_0(m_2))  |\mathcal{M}_1(m_1)|^2$ & $-\frac{2}{\sqrt{3}} \cos^2\theta_1 \cos\theta_2$ \\
    9 & $\frac{1}{\sqrt{2}} \frac{1}{\Gamma_L} {|A_s^-|} {|A_{\parallel}|} \Real( e^{i({\delta_s^-} - {\delta_{\parallel}})} \mathcal{M}_1^*(m_2) \mathcal{M}_0(m_2))  |\mathcal{M}_1(m_1)|^2$ & $-\frac{1}{\sqrt{6}} \sin 2\theta_1 \sin\theta_2 \cos\varphi$ \\
    10 & $\frac{1}{\sqrt{2}} \frac{1}{\Gamma_H} {|A_s^+|} {|A_{\perp}|} \Imag(e^{i( {\delta_{\perp}} - {\delta_s^+} )} \mathcal{M}_0^*(m_2) \mathcal{M}_0(m_2) ) |\mathcal{M}_1(m_1)|^2$ & $\frac{1}{\sqrt{6}} \sin 2\theta_1 \sin\theta_2 \sin\varphi$ \\
    11 & $\frac{1}{\sqrt{2}} \frac{1}{\Gamma_L} {|A_s^-|} {|A_{ss}|} \Real( e^{i({\delta_s^-} - {\delta_{ss}})} \mathcal{M}_0^*(m_1) \mathcal{M}_1(m_1))  |\mathcal{M}_0(m_2)|^2$ & $\frac{2}{3\sqrt{3}} \cos\theta_1$ \\
    \hline
    12 & $\frac{1}{2}  (\frac{{|A_s^+|}^2}{\Gamma_H} + \frac{{|A_s^-|}^2}{\Gamma_L} ) |\mathcal{M}_0(m_1)|^2 |\mathcal{M}_1(m_2)|^2$ & $\frac{1}{3} \cos^2\theta_2$ \\
    13 & $- \frac{1}{\sqrt{2}} \frac{1}{\Gamma_L} {|A_s^-|} {|A_{0}|}  \Real( e^{i{\delta_s^-}} \mathcal{M}_1^*(m_1) \mathcal{M}_0(m_1))   |\mathcal{M}_1(m_2)|^2$ & $\frac{2}{\sqrt{3}} \cos\theta_1 \cos^2\theta_2$ \\
    14 & $- \frac{1}{\sqrt{2}} \frac{1}{\Gamma_L} {|A_s^-|} {|A_{\parallel}|} \Real( e^{i({\delta_s^-} -{ \delta_{\parallel}})} \mathcal{M}_1^*(m_1) \mathcal{M}_0(m_1))  |\mathcal{M}_1(m_2)|^2$ & $\frac{1}{\sqrt{6}} \sin\theta_1 \sin 2\theta_2 \cos\varphi$ \\
    15 & $\frac{1}{\sqrt{2}} \frac{1}{\Gamma_H} {|A_s^+|} {|A_{\perp}|}  \Imag(e^{i( {\delta_{\perp}} - {\delta_s^+} )} \mathcal{M}_0^*(m_1) \mathcal{M}_0(m_1) ) |\mathcal{M}_1(m_2)|^2$ & $-\frac{1}{\sqrt{6}} \sin\theta_1 \sin 2\theta_2 \sin\varphi$ \\
    16 & $- \frac{1}{\sqrt{2}} \frac{1}{\Gamma_L} {|A_s^-|} {|A_{ss}|} \Real( e^{i({\delta_s^-} - {\delta_{ss}})} \mathcal{M}_0^*(m_2) \mathcal{M}_1(m_2))  |\mathcal{M}_0(m_1)|^2$ & $-\frac{2}{3\sqrt{3}} \cos\theta_2$ \\
    \hline
    17 &  $ (\frac{{|A_s^+|}^2}{\Gamma_H} - \frac{{|A_s^-|}^2}{\Gamma_L} ) \Real(\mathcal{M}^*_1(m_1)\mathcal{M}^*_0(m_2) \mathcal{M}_0(m_1)\mathcal{M}_1(m_2) ) $ & $-\frac{1}{3} \cos\theta_1 \cos\theta_2$  \\
    \hline
    18 & $\frac{1}{\Gamma_L} {|A_{ss}|}^2 |\mathcal{M}_0(m_1)|^2 |\mathcal{M}_0(m_2)|^2 $ & $\frac{1}{9}$ \\
    19 & $\frac{1}{\Gamma_L} {|A_{ss}|} {|A_{0}|} \Real(e^{i{\delta_{ss}}}  \mathcal{M}_1^*(m_1) \mathcal{M}_1^*(m_2) \mathcal{M}_0(m_1) \mathcal{M}_0(m_2) )$ & $-\frac{2}{3} \cos\theta_1 \cos\theta_2$ \\
    20 & $\frac{1}{\Gamma_L} {|A_{ss}|} {|A_{\parallel}|} \Real(e^{i({\delta_{ss}}-{\delta_{\parallel}})}  \mathcal{M}_1^*(m_1) \mathcal{M}_1^*(m_2) \mathcal{M}_0(m_1) \mathcal{M}_0(m_2) )$ & $-\frac{\sqrt{2}}{3} \sin\theta_1 \sin\theta_2 \cos\varphi$ \\
    21 & $0$ & $\frac{\sqrt{2}}{3} \sin\theta_1 \sin\theta_2 \sin\varphi$ \\
    \hline
  \end{tabular}
\end{table}

\afterpage{\clearpage}

The dependence of each amplitude on the invariant mass of the $K^+\pi^-$ and $K^-\pi^+$ pairs is given by the 
propagators $\mathcal{M}_{J}(m) \propto \mathcal{R}_J(m) \times m/q$, where $q$ is the momentum of each meson 
in the rest frame of the $K\pi$ pair
\begin{equation}
\label{eq:momentum}
q = \frac{\sqrt{(m^2-(M_{\pi}+M_{K})^2 ) (m^2 -(M_{\pi}-M_{K})^2}) }{2m}.
\end{equation}

The \pwave propagator, $J=1$, is parameterised using a spin-1 relativistic Breit-Wigner
resonance function
\begin{equation}
  \mathcal{R}_1(m) = \frac{M_1 \Gamma_1(m)}{(M_1^2 - m^2) - iM_1\Gamma_1(m) }.
\label{eq:propag}
\end{equation}
\noindent The  mass-dependent width is given by
\begin{equation}
  \Gamma_1(m) = \Gamma_1 \frac{M_1}{m} \frac{1+r^2 q_1^2}{1 + r^2 q^2} \left( \frac{q}{q_1}\right)^3,
\end{equation}
\noindent where $M_1$ and $\Gamma_1$ are the $\Kstarz(892)$ resonance mass and width, $r$ is the interaction radius
and $q_1$ corresponds to \eqref{eq:momentum} evaluated at the resonance position ($M_1$).

To describe the \swave propagator, $\mathcal{M}_0(m)$, the LASS parameterisation~\cite{LASS} is used, which is an effective-range elastic scattering 
amplitude, interfering with the $\Kstars(1430)$ resonance,
\begin{equation}
\label{eq:bwKpiSwave}
 \mathcal{R}_0(m) \propto \frac{1}{\cot\delta_{\beta} - i} + e^{2i\delta_{\beta}}\frac{M_0 \Gamma_0(m)}{M_0^2 - m^2 - i M_0 \Gamma_0(m) },
\end{equation}
\noindent where
\begin{equation}
\label{eq:Gamma0}
 \Gamma_0(m) = \Gamma_0 \frac{M_0}{m} \left ( \frac{q}{q_0} \right ),
\end{equation}
\noindent and the non-resonant component is described as
\begin{equation}
\label{eq:DRDBequation}
  \cot\delta_{\beta} = \frac{1}{a q} + \frac{1}{2}bq.
\end{equation}
The values of the mass propagator parameters, including the resonance masses and widths, $M_J$ and $\Gamma_J$, and the
the scattering length ($a$) and effective range ($b$), are summarized in \tabref{table:inputSwave}. 
Other shapes modelling the \swave propagator, including an explicit Breit-Wigner contribution for the 
$\Kstars(800)$ resonance, are considered in the systematic uncertainties.



The normalisation of the mass propagators 
\begin{equation}
	\int |\mathcal{M}_0|^2 \mathrm{d}m = \int |\mathcal{M}_1|^2 \mathrm{d}m =1
\end{equation}

\noindent in the mass range considered, together with the normalisation condition 
\begin{equation}
  |A_0|^2+|A_{\parallel}|^2+|A_{\perp}|^2+|A_s^{+}|^2+|A_s^{-}|^2 +|A_{ss}|^2 = 1  ,
\end{equation}
guarantees the definition of the squared amplitudes as fractions of different partial waves.
The polarisation fractions for the vector mode, \BsKstKst, are defined as 
\begin{equation}
  f_{L,\parallel,\perp} = \frac{|A_{0,\parallel,\perp}|^2}{|A_0|^2+|A_{\parallel}|^2+|A_{\perp}|^2} \, .
\end{equation}

\noindent The overall phase of the propagators is defined such that
\begin{equation}
  \arg[\mathcal{M}_0(M_1)] = \arg[\mathcal{M}_1(M_1)] = 0  
\end{equation}
and the convention $\delta_0 \equiv \arg(A_0) = 0$ is adopted. Therefore $\delta_{\parallel}$, $\delta_{\perp}$, 
$\delta_s^-$, $\delta_s^+$ and $\delta_{ss}$ are defined as the phase difference between the 
corresponding amplitude and $A_0$ at the \Kstarz mass pole. As a consequence of the lack of initial \Bs or \Bsb flavour information,
the phases $\delta_{\perp}$ and $\delta_{s}^+$ can not be measured independently, and only 
their difference is accessible to this analysis.

\begin{table}[t]
\begin{center}
\caption{Parameters of the mass propagators used in the fit.}
\begin{tabular}{l|c|c}
   & $(K\pi)_0^{*0}$  & $K^{*}(892)^0$  \\
       & $J=0$~\cite{LASS,bdphikst_babar}  & $J=1$~\cite{PDG}  \\
  \hline
   $M_J$ (\mevcc)  & $1435 \pm 5 \pm \phantom{1}5$ & $895.81 \pm 0.19$ \\
   $\Gamma_J$ (\mevcc)  & $\phantom{1}279 \pm 6 \pm 21$ & $\phantom{0}47.4\phantom{1} \pm 0.6\phantom{0}$  \\
   $r$ ($\gev^{-1}$)    & - & $\phantom{00}3.0\phantom{1} \pm 0.5\phantom{0}$ \\
   $a$ ($\gev^{-1}$)    & $1.95 \pm 0.09 \pm 0.06$ & - \\
   $b$ ($\gev^{-1}$)    & $1.76 \pm 0.36 \pm 0.67$ & -  \\
\end{tabular}
\label{table:inputSwave}
\end{center}
\end{table}

\section{The LHCb detector}
\label{sec:detector}

The \lhcb detector~\cite{Alves:2008zz,LHCb-DP-2014-002} is a single-arm forward
spectrometer covering the \mbox{pseudorapidity} range $2<\eta <5$,
designed for the study of particles containing \bquark or \cquark
quarks. The detector includes a high-precision tracking system
consisting of a silicon-strip vertex detector surrounding the $pp$
interaction region, a large-area silicon-strip detector located
upstream of a dipole magnet with a bending power of about
$4{\rm\,Tm}$, and three stations of silicon-strip detectors and straw
drift tubes placed downstream of the magnet.
The tracking system provides a measurement of momentum, \ptot, of charged particles with
a relative uncertainty that varies from 0.5\% at low momentum to 1.0\% at 200\gevc.
The minimum distance of a track to a primary vertex, the impact parameter (IP), is measured with a resolution of $(15+29/\pt)\mum$,
where \pt is the component of the momentum transverse to the beam, in\,\gevc.
Different types of charged hadrons are distinguished using information
from two ring-imaging Cherenkov detectors. 
Photons, electrons and hadrons are identified by a calorimeter system consisting of
scintillating-pad and preshower detectors, an electromagnetic
calorimeter and a hadronic calorimeter. Muons are identified by a
system composed of alternating layers of iron and multiwire
proportional chambers.
The online event selection is performed by a trigger, 
which consists of a hardware stage, based on information from the calorimeter and muon
systems, followed by a software stage, which applies a full event
reconstruction.

In the analysis presented here, all
hardware triggers are used.
The software trigger requires a multi-track
secondary vertex with a significant displacement from the primary
$pp$ interaction vertices~(PVs). At least one charged particle
must have a transverse momentum $\pt > 1.7\gevc$ and be
inconsistent with originating from a PV.
A multivariate algorithm~\cite{BBDT} identifies secondary vertices consistent with the decay
of a \bquark hadron. 

Simulated \BsKstKst events are used to characterise the detector response to signal events. 
In the simulation, $pp$ collisions are generated using
\pythia~\cite{Sjostrand:2006za} 
with a specific \lhcb
configuration~\cite{LHCb-PROC-2010-056}.  Decays of hadronic particles
are described by \evtgen~\cite{Lange:2001uf}, in which final-state
radiation is generated using \photos~\cite{Golonka:2005pn}. The
interaction of the generated particles with the detector and its
response are implemented using the \geant
toolkit~\cite{Allison:2006ve, *Agostinelli:2002hh} as described in
Ref.~\cite{LHCb-PROC-2011-006}.

\section{Event selection and signal yield}
\label{sec:selection}

The event selection is similar to that used in the previous analysis~\cite{KstKstPaper}. 
\Kstarz candidates are formed from two high-quality oppositely charged tracks identified as a kaon and pion, respectively.
They are selected to have $\pt>500$\mevc and to be displaced from any PV.
The $K^+\pi^-$ and $K^-\pi^+$ pairs are required to have invariant mass within $\pm 150$\mevcc of the known \Kstarz mass,
which corresponds to 74\% of the total phase-space for \BsKstKst, 
and $\pt>900$\mevc. Each \Bs candidate is constructed by combining a \Kstarz and \Kstarzb, requiring the four tracks 
to form a good vertex well-separated from any PV. The \Bs candidate invariant mass is restricted to be within 
the interval $[5100,5866]$\mevcc and its momentum vector is required to point towards one PV.

In order to further discriminate the \BsKstKst signal from the combinatorial background, different properties of the
decay are combined into a multivariate discriminator~\cite{GL}.
The variables combined in the discriminator are the \Bs candidate
IP with respect to the associated PV, its lifetime and \pt, the minimum \chisqip of the four daughter tracks (defined as
the difference between the $\chi^2$ of a PV formed with and without the particle in question) with respect to 
the same PV and the distance of closest approach between the two \Kstarz candidates. 
The discriminator is trained using simulated \BsKstKst events for signal and a small data sample excluded
from the rest of the analysis as background. The optimal
discriminator requirement is determined by maximising the figure of merit $N_S/\sqrt{N_S+N_B}$ in a test sample
containing signal ($S$) and background ($B$) events of the same nature as those used in the training sample.

Differences in the log-likelihood for various particle identification hypotheses ($\DLL_{a-b}$) 
are used to minimise the contamination from specific \B decays. Contributions from \BdRhoKst and \BdPhiKst modes are reduced by the $\DLL_{K-\pi}$
requirements of kaons and pions. A small contamination from \Lambdabppikst decays is observed and suppressed with
$\DLL_{{p}-K}$ requirements. 

\begin{figure}
  \centering
  \includegraphics[width=0.8\textwidth]{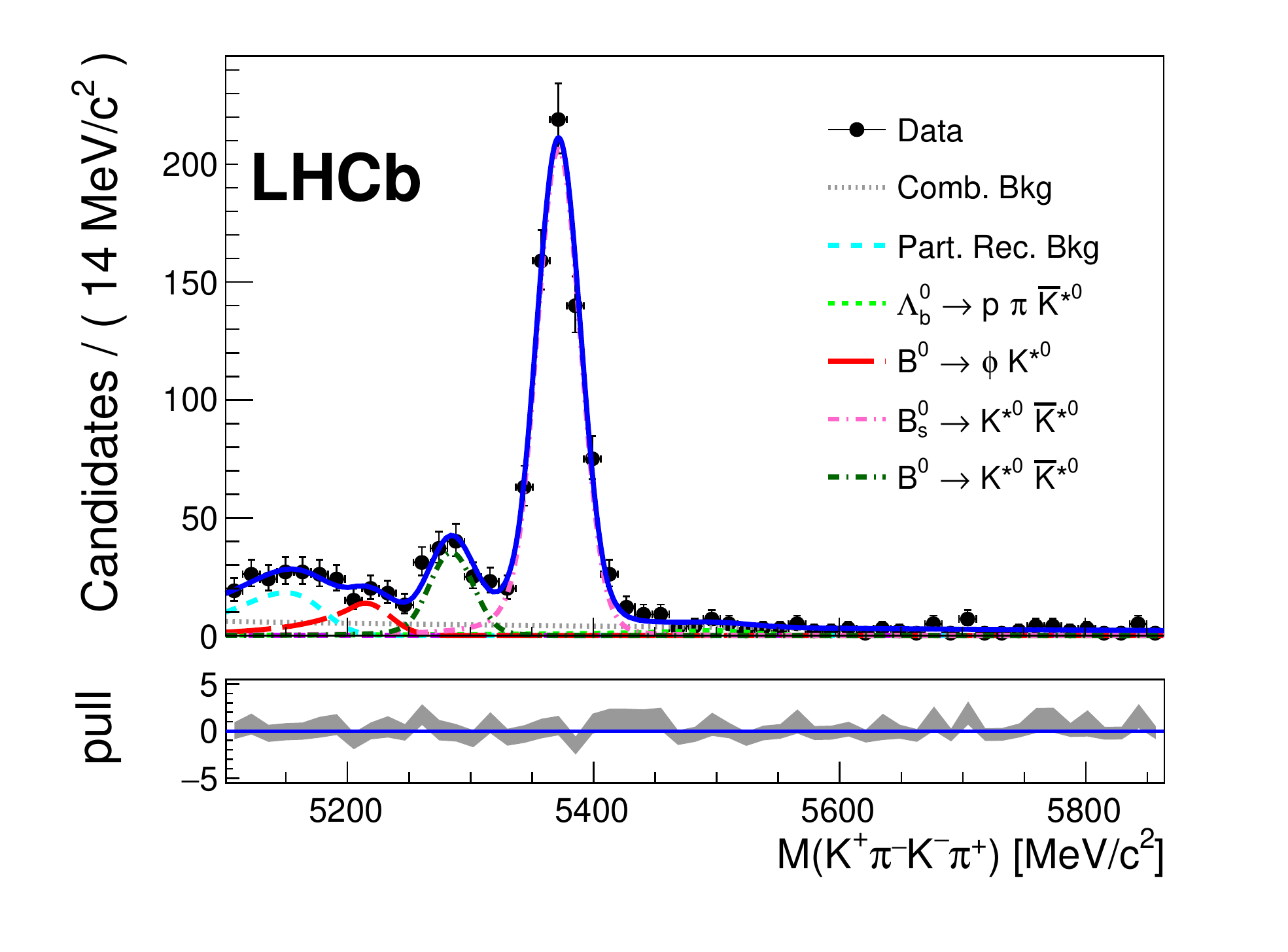}
  \caption{Invariant mass distribution for selected $K^+\pi^- K^- \pi^+$ candidates. The (blue) solid line is the result of the 
    fit explained in the text. The \Bs and \Bd signal peaks are shown as as dashed-dotted lines (pink and dark green, respectively). 
	The various peaking background components
	are represented as dotted lines: (red) \BdPhiKst,  (green) \Lambdabppikst
	and (light blue) partially reconstructed decays. The (grey) dotted line is the
	combinatorial background component. The normalised residual (pull) is shown below.
}
\label{fig:massfit}
\end{figure}

An extended unbinned maximum likelihood fit to the mass spectrum of the selected \BsKpiKpi
candidates is performed. The signal is modelled by a sum of two Crystal Ball distributions~\cite{crystalball} that share common mean and 
width. The same distribution is used to describe the \Bd decay into the same final state. Components for \BdPhiKst and \Lambdabppikst decays are included
in the fit with shapes extracted from simulated events. The contribution from \BdRhoKst decays is estimated to be negligible 
from simulation studies. Finally, partially reconstructed \B decays are parameterised 
using an ARGUS distribution~\cite{argus} and the remaining combinatorial background is modelled using an exponential function.
The fit result is shown in \figref{fig:massfit}. A total of
$697 \pm 31$ \BsKpiKpi decays is obtained. 

\subsection{Acceptance properties}
\label{sec:acceptance}

Effects introduced in data due to the geometry of the detector and to the selection requirements 
need to be taken into account in the measurement.

The study of simulated \BsKstKst events shows that the detection and selection efficiency is not 
uniform as a function of the decay angles $\theta_1$ and $\theta_2$, but has no dependence, at the level 
of precision needed for this analysis, on $\varphi$ and on the invariant mass of the two $K\pi$ pairs, $m_1$ and $m_2$. 
The acceptance decreases as $\cos\theta_i$ approaches 1. This feature is mainly 
induced by the requirement on the minimum \pt of the daughter pions. This effect is modelled by a 
two-dimensional function in $\cos\theta_1$ and $\cos\theta_2$, which is extracted from simulation. 

Since the trigger system uses the \pt of the charged particles, the acceptance effect is different 
for events where signal tracks were involved in the trigger decision (called trigger-on-signal or TOS throughout) 
and those where the trigger decision was made using information from the rest of the event (non-TOS). The data set 
is split according to these two categories and a different acceptance correction is applied to each subset.

\section{Triple product and direct $\mathbf{CP}$ asymmetries}
\label{sec:tpa}

Triple products and direct \CP asymmetries are calculated for \BsKpiKpi using Eqs.~(\ref{formula:TP_1}) and~(\ref{formula:TP_2}), after time integration, and Eqs.~(\ref{formula:TP_3})--(\ref{formula:DCPA_4})
from those candidates with a four-body invariant mass within $\pm 30$ \mevcc of the known \Bs mass. 
The background in this interval, which is purely combinatorial, is subtracted according to the fraction calculated
from the result of the invariant mass fit, $f_{\rm bkg} = (3.44 \pm 0.34)\%$. The angular distributions of the background are
extracted from the upper mass sideband, defined by $M(K^+ \pi^- K^- \pi^+)>5550$ \mevcc. Acceptance effects
are then corrected in the signal angular distributions. The measured asymmetries are listed in \tabref{tab:TP_AD}.
From the definitions given in \secref{sec:tpa_def}, correlations of the order of 5\% are expected among these
asymmetries, with the exception of $A_D^1$ and $A_D^3$ where the correlation is calculated to be close to 90\%. 

\begin{table}
  \caption{Triple product and direct \CP asymmetries measured in this analysis. The first uncertainties are statistical and the second systematic.}
  \centering
  \begin{tabular}{cc}
    \hline
    Asymmetry & Value \\
    \hline
    $A_{T}^1$  & $\phantom{-}0.003$ $\pm$ 0.041 $\pm$ 0.009  \\   
    $A_{T}^2$  & $\phantom{-}0.009$ $\pm$ 0.041 $\pm$ 0.009  \\  
    $A_{T}^3$  & $\phantom{-}0.019$ $\pm$ 0.041 $\pm$ 0.008  \\ 
    $A_{T}^4$  & $-0.040$ $\pm$ 0.041 $\pm$ 0.008  \\ 
    \hline
    $A_{D}^1$  & $-0.061$ $\pm$ 0.041 $\pm$ 0.012  \\ 
    $A_{D}^2$  & $\phantom{-}0.081$ $\pm$ 0.041 $\pm$ 0.008  \\ 
    $A_{D}^3$  & $-0.079$ $\pm$ 0.041 $\pm$ 0.023  \\
    $A_{D}^4$  & $-0.081$ $\pm$ 0.041 $\pm$ 0.010  \\ 
    \hline
  \end{tabular}
  \label{tab:TP_AD}
\end{table}

The main systematic uncertainty in these measurements is associated to the angular acceptance correction. 
Discrepancies in the \pt spectra and the particle identification efficiencies between data and simulation are 
used to modify the acceptance function obtained from simulation. Systematic uncertainties are
determined from the variation in the measured asymmetries when this modified acceptance is used. Systematic
effects are found to be larger in case of the four direct \CP asymmetries, 
in particular for $A_{D}^{3}$, which has a strong dependence on $\cos\theta_{1,2}$. 
In addition, the lifetime-biasing selection criteria have a slightly different effect on the various amplitudes, 
which correspond to decays with different effective lifetimes, due to the width difference between \Bs mass eigenstates~\cite{deBruyn,LHCb_DeltaGamma}. This could 
induce a bias in the measured TP and direct \CP asymmetries. A set of simulated experiments is performed to estimate the impact
of the lifetime acceptance in the eight quantities. The observed deviations are small and are used to assign a systematic uncertainty.
Finally, the effect of the uncertainty in the background contribution is estimated by changing the background 
fraction and the parameters in the background model within their statistical uncertainty and recalculating the asymmetries.

\section{Angular analysis}
\label{sec:angular}

The magnitudes and phases of the various amplitudes contributing to the \BsKpiKpi decay 
are determined using a five-dimensional fit 
to the three helicity angles ($\Omega$) and to the invariant mass of the 
two $K\pion$ pairs ($m_1, m_2$) of all candidates with a four-body 
invariant mass $|M(K^+, \pi^-, K^-, \pi^+) - m_{\Bs}| < 30 \mevcc$.

The model used to describe the distribution in these five variables is given by
\begin{equation}
  \mathcal{F}(\Omega, m_1,m_2) = (1-f_{\rm bkg}) F(\Omega,m_1,m_2) \times \varepsilon(\Omega) + f_{\rm bkg} F_{\rm bkg}(\Omega,m_1,m_2)  ,
\end{equation}
where $F(\Omega,m_1,m_2)$ is the probability density function in \eqref{eq:rate_timedep}, $\varepsilon(\Omega)$ is the acceptance
function modelling the effects introduced by reconstruction, selection and trigger reported in \secref{sec:acceptance}, and $F_{\rm bkg}(\Omega,m_1,m_2)$ describes
the distribution of the background extracted from the upper mass sideband. 
The background fraction, $f_{\rm bkg}$, 
is obtained from the result of the fit to the invariant mass of the \Bs candidates.

\begin{figure}[p]
  \centering
  \includegraphics[width=0.32\textwidth]{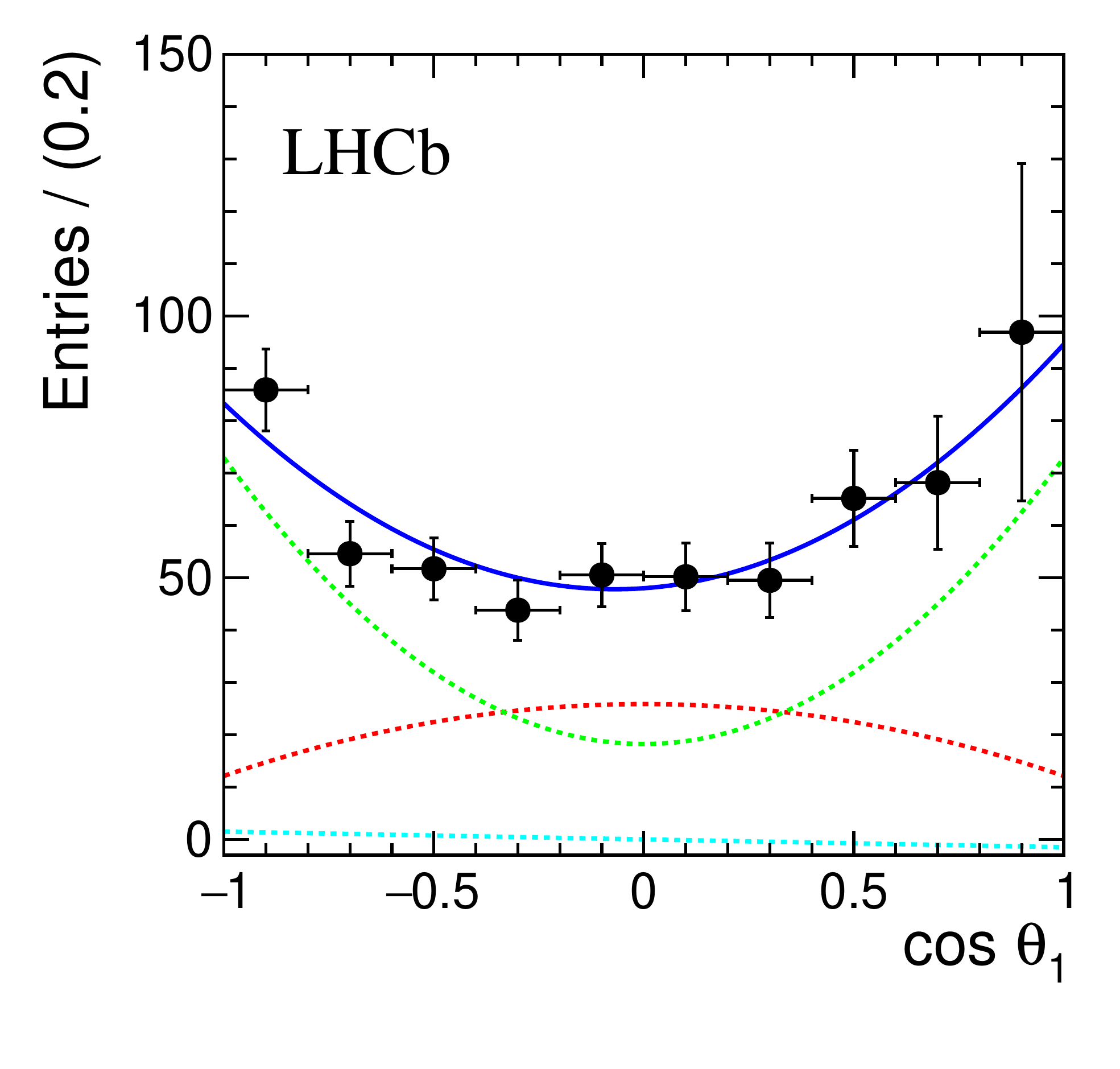}
  \hfill
  \includegraphics[width=0.32\textwidth]{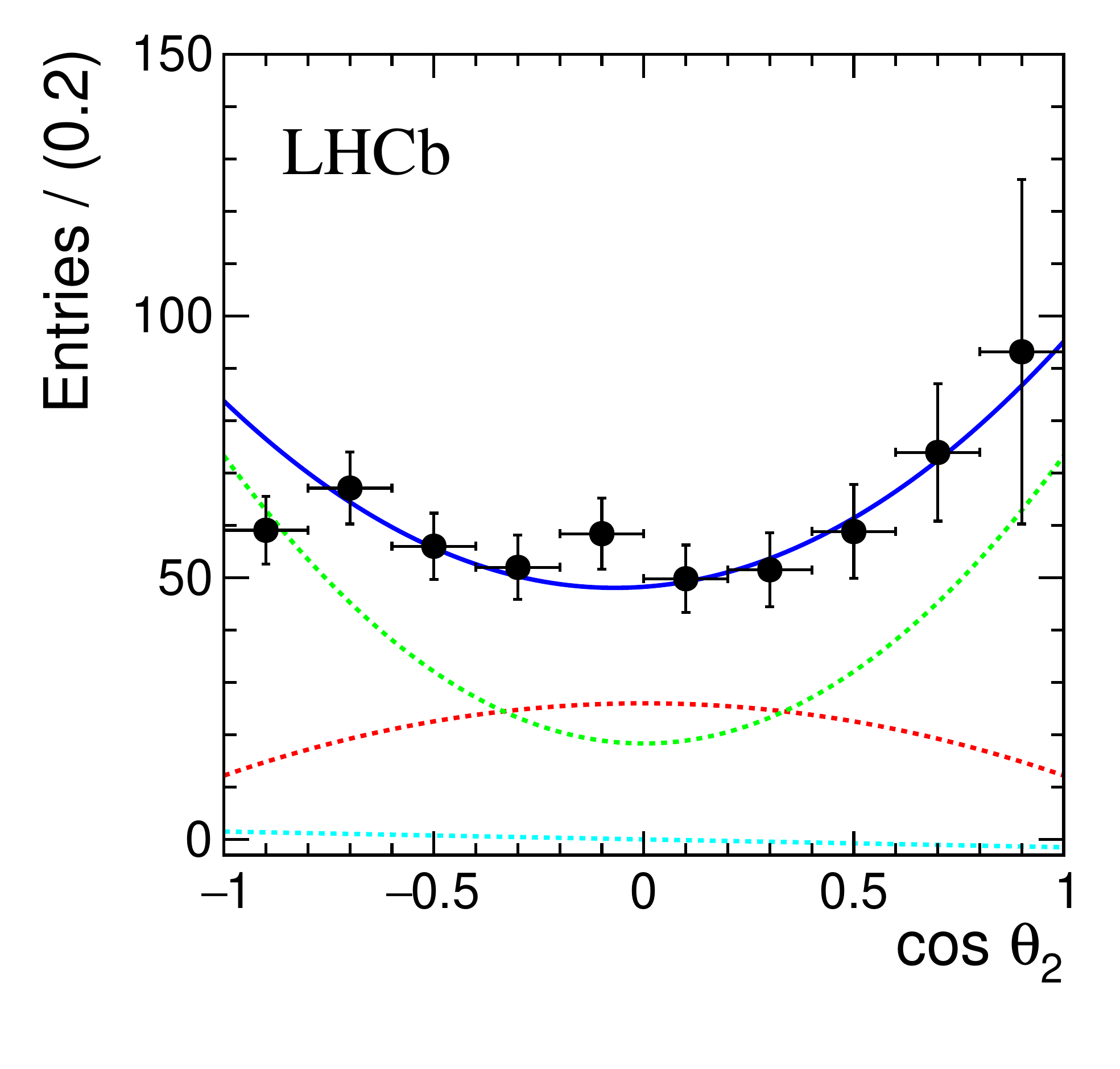}
  \hfill
  \includegraphics[width=0.32\textwidth]{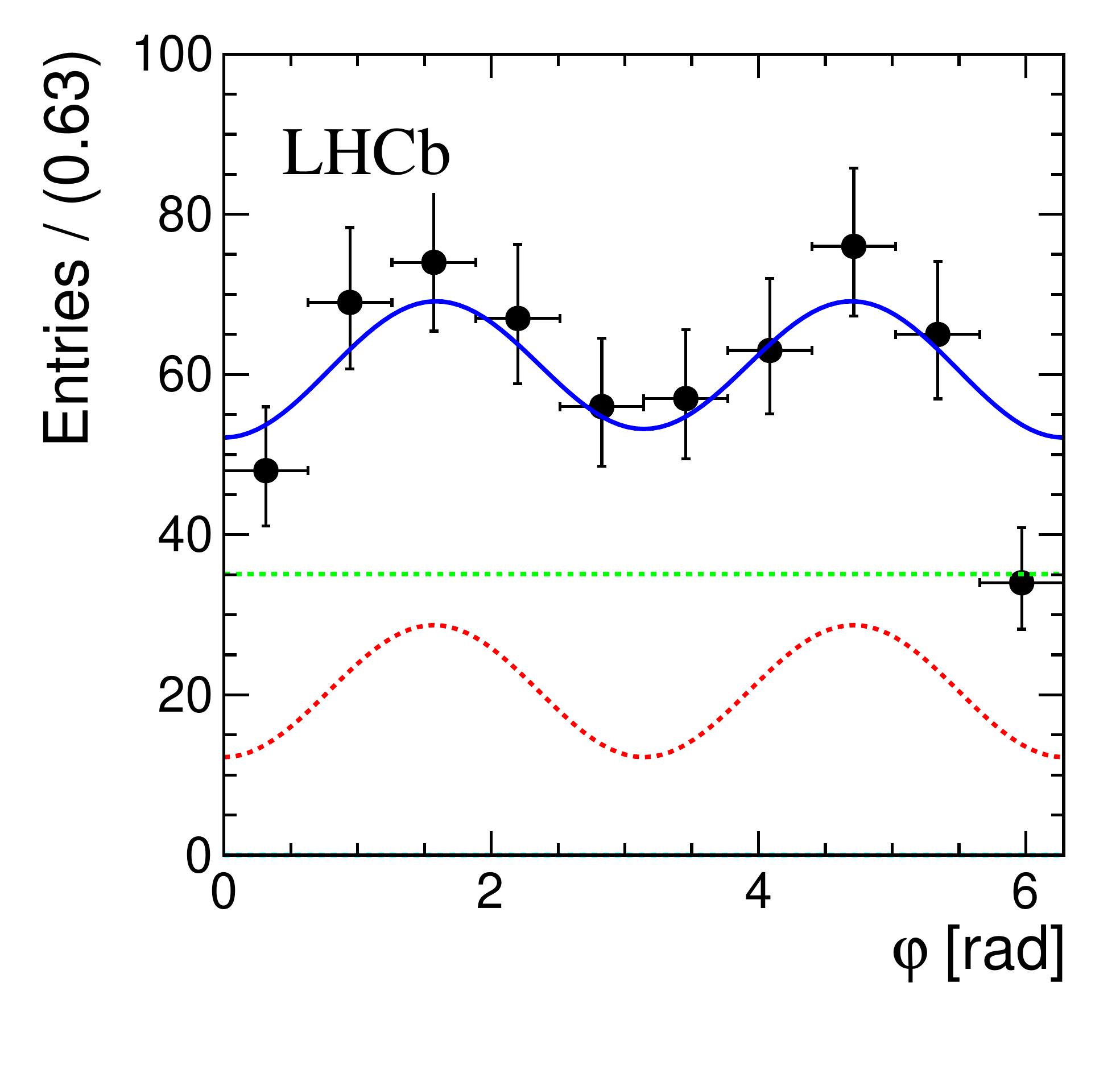}
  \caption{Results of the simultaneous fit to \BsKpiKpi candidates (blue solid line) in the three helicity angles.
           The dots represent the data after background subtraction and acceptance correction.
		   The red dashed line is the \pwave component, the green dashed line is the \swave component and the light-blue dashed 
           line represents the $\mathcal{A}^+_s \mathcal{A}_0$ interference term. 
		   }
  \label{fig:fit_kstkst}
\end{figure}

\begin{figure}[p]
  \centering
  \includegraphics[width=0.32\textwidth]{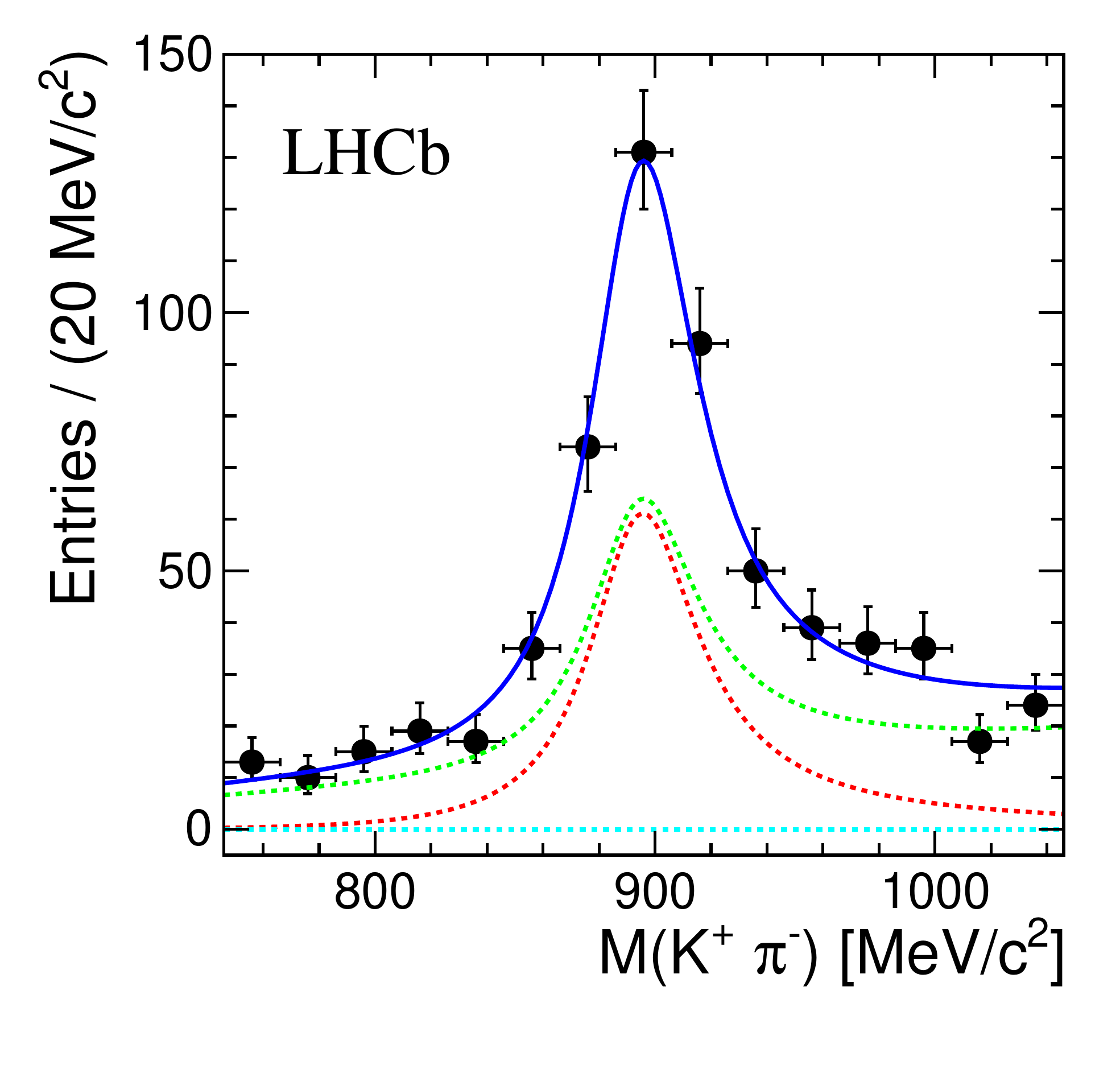}
  \hspace{1mm}
  \includegraphics[width=0.32\textwidth]{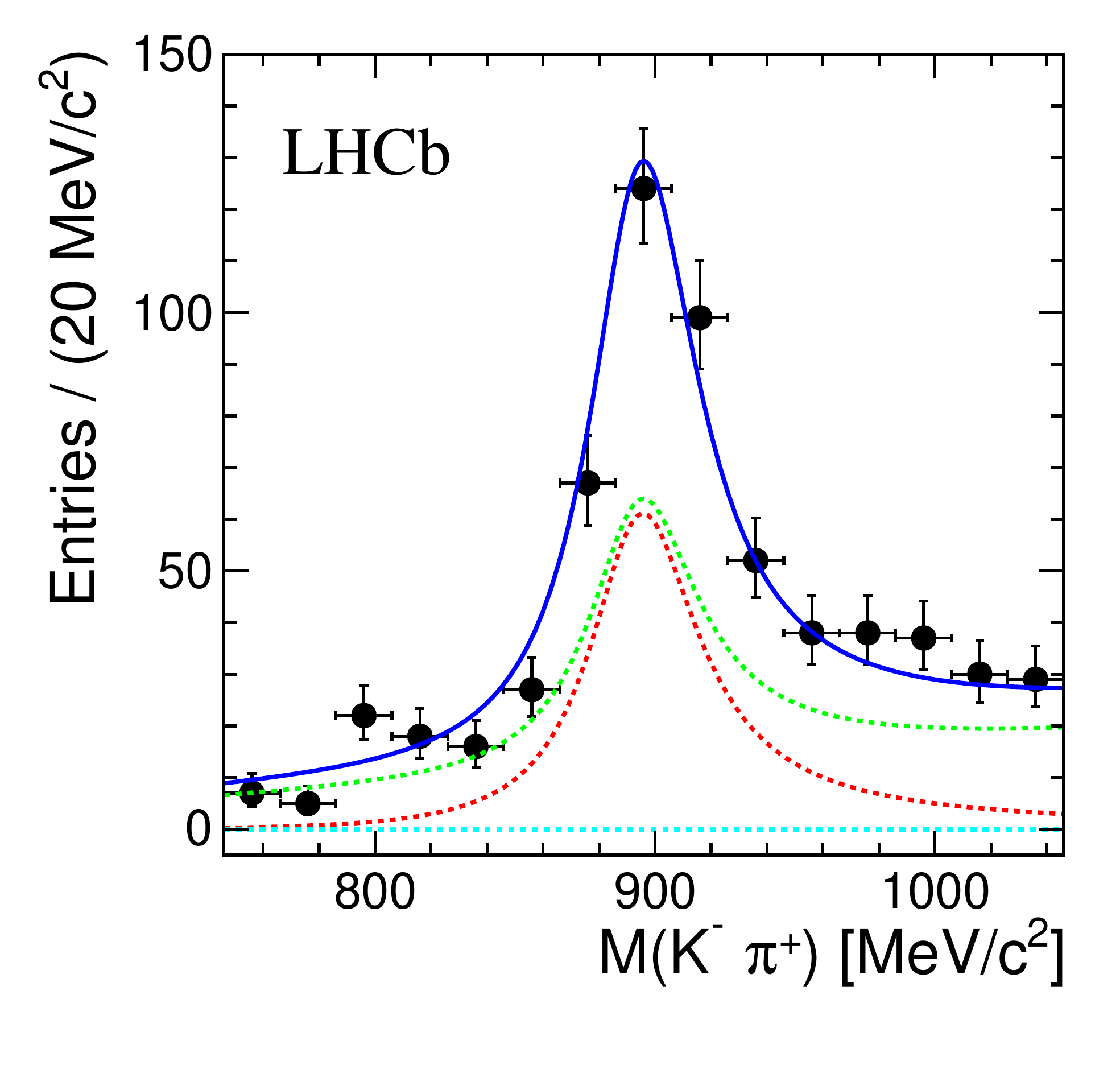}

  \vspace{-0.3cm}
  \includegraphics[width=0.32\textwidth]{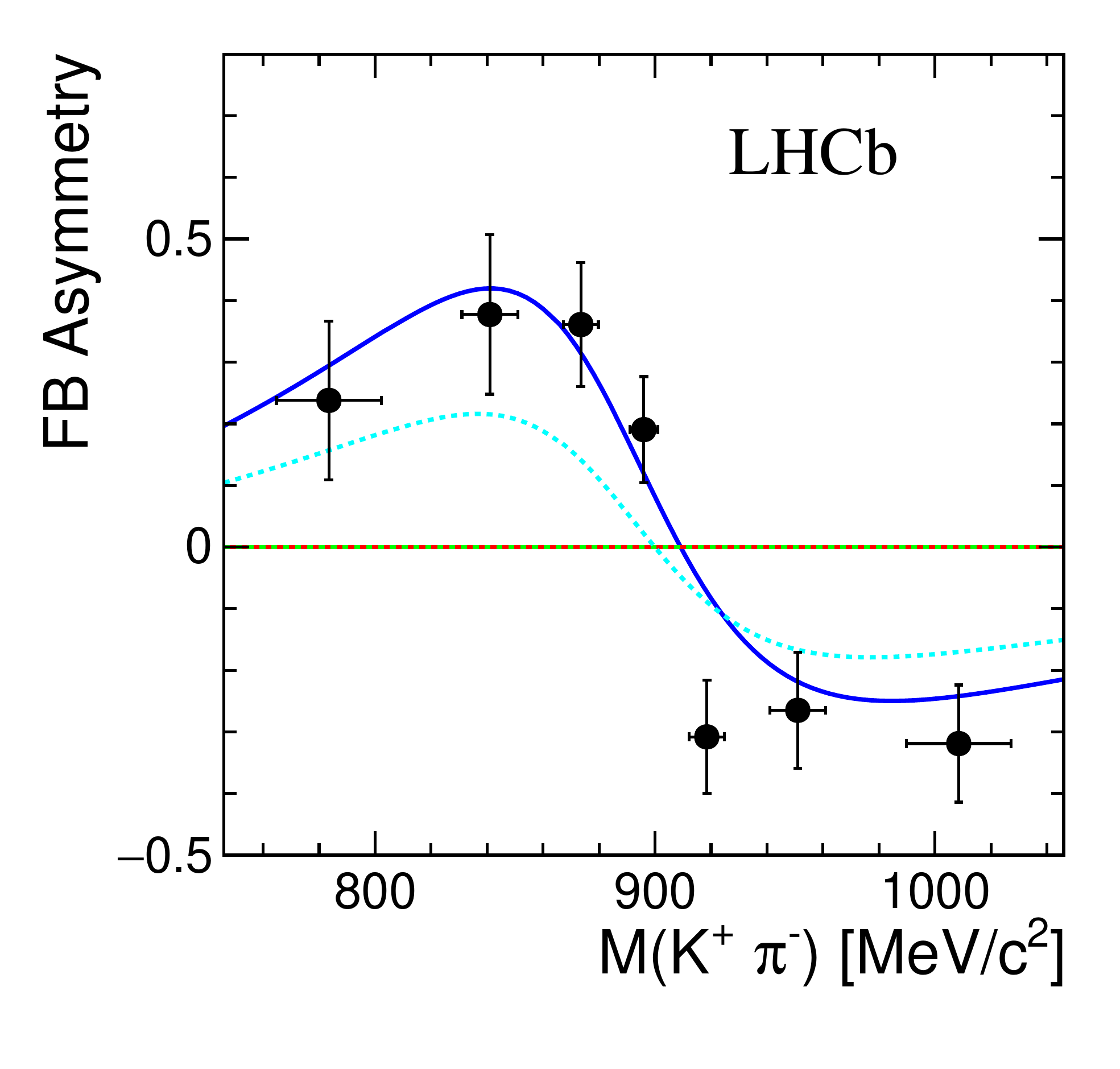}
  \hspace{1mm}
  \includegraphics[width=0.32\textwidth]{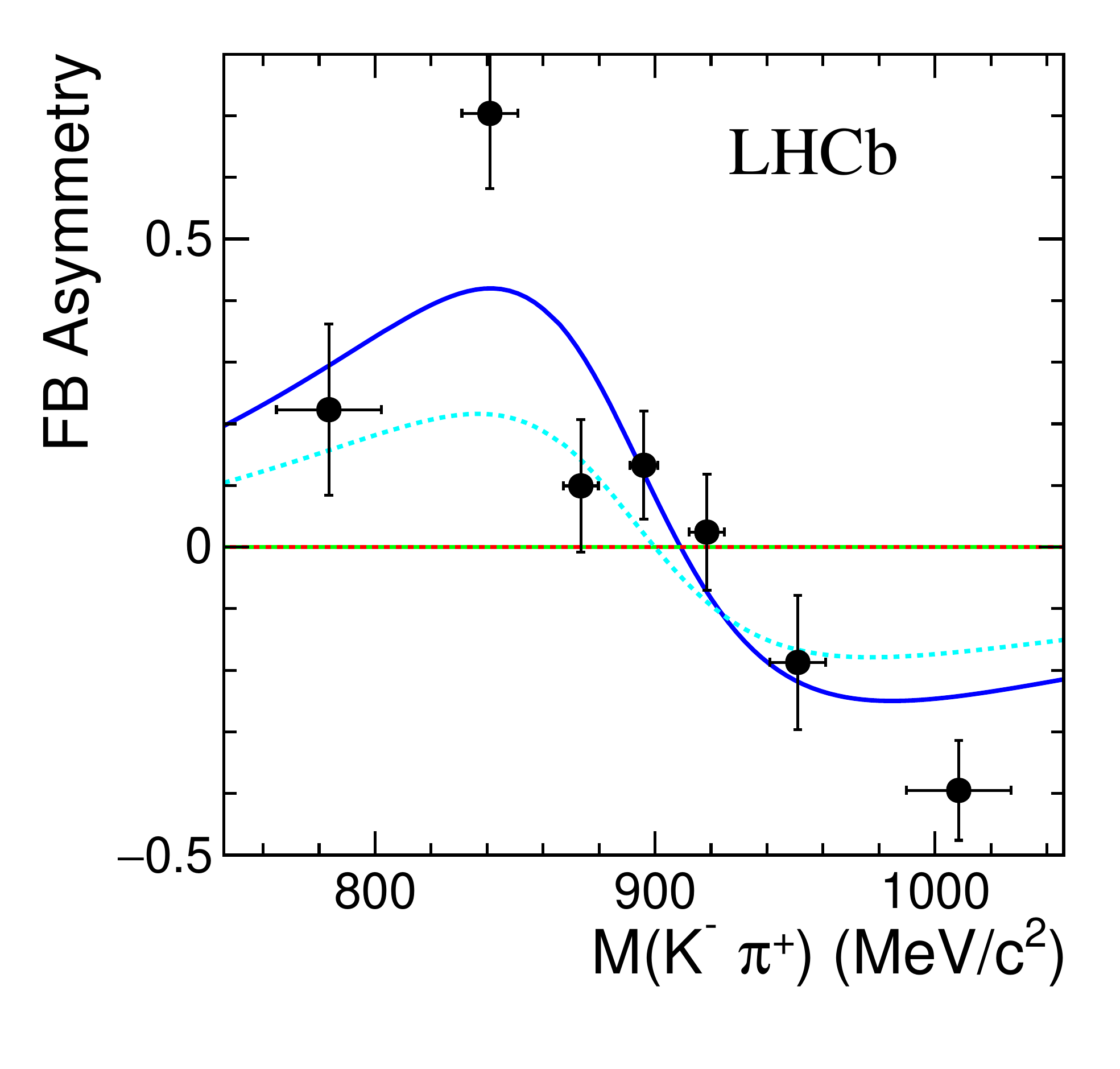}
  \caption{Projections of the model fitted to \BsKpiKpi candidates (blue solid line) in the (top) invariant mass of $K\pi$ pairs 
    and (bottom) $\cos\theta$ asymmetries as functions of $K\pi$ mass ($m_1$, $m_2$). The dots represent the data after 
	background subtraction and acceptance correction.
	The red dashed line is the \pwave component, the green dashed line is the \swave component and the light-blue dashed 
    line represents the ${A}^+_s {A}_0$ interference term. }
  \label{fig:fit_kstkst_2}
\end{figure}

Using this model, an unbinned maximun likelihood fit is performed simultaneously for TOS and non-TOS \BsKpiKpi 
candidates, where only the acceptance function and the background fraction are different between the two samples. 
The results of the fit are summarised in \tabref{tab:fit_kstkst}.
Figures \ref{fig:fit_kstkst} and \ref{fig:fit_kstkst_2} show the angular and $K\pi$ mass projections of the multi-dimensional distributions.
To quantitatively demonstrate the interference between the different partial waves, a forward-backward asymmetry 
is defined for \Kstarz meson as $A_{\rm FB} = (N_{\rm F}-N_{\rm B})/(N_{\rm F}+N_{\rm B})$, where $N_{\rm F}$ ($N_{\rm B}$) is the number of 
$K^+$ mesons emitted with positive (negative) $\cos\theta_1$, and analogously for the \Kstarzb meson. 
Their evolution with the $K\pi$ invariant mass is shown in \figref{fig:fit_kstkst_2}, as an 
additonal projection of the fit result. 
According to \eqref{eq:rate_timedep}
these asymmetries are proportional to the interference term between $A_s^-$ and $A_0$.

\begin{figure}[p]
  \centering
  \begin{minipage}{0.4\textwidth}
    \includegraphics[width=\textwidth]{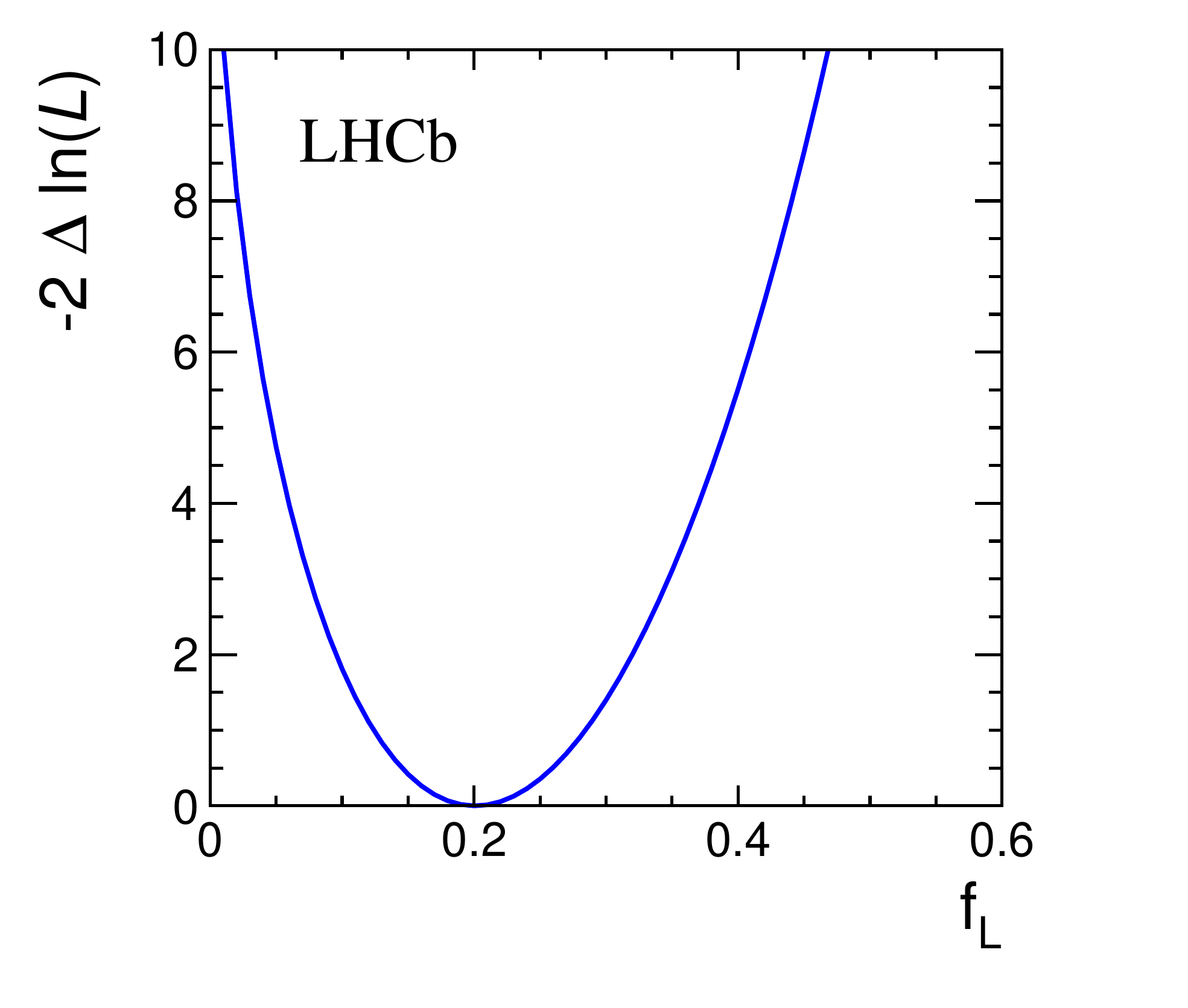}
  \end{minipage}
  \begin{minipage}{0.4\textwidth}
  \includegraphics[width=\textwidth]{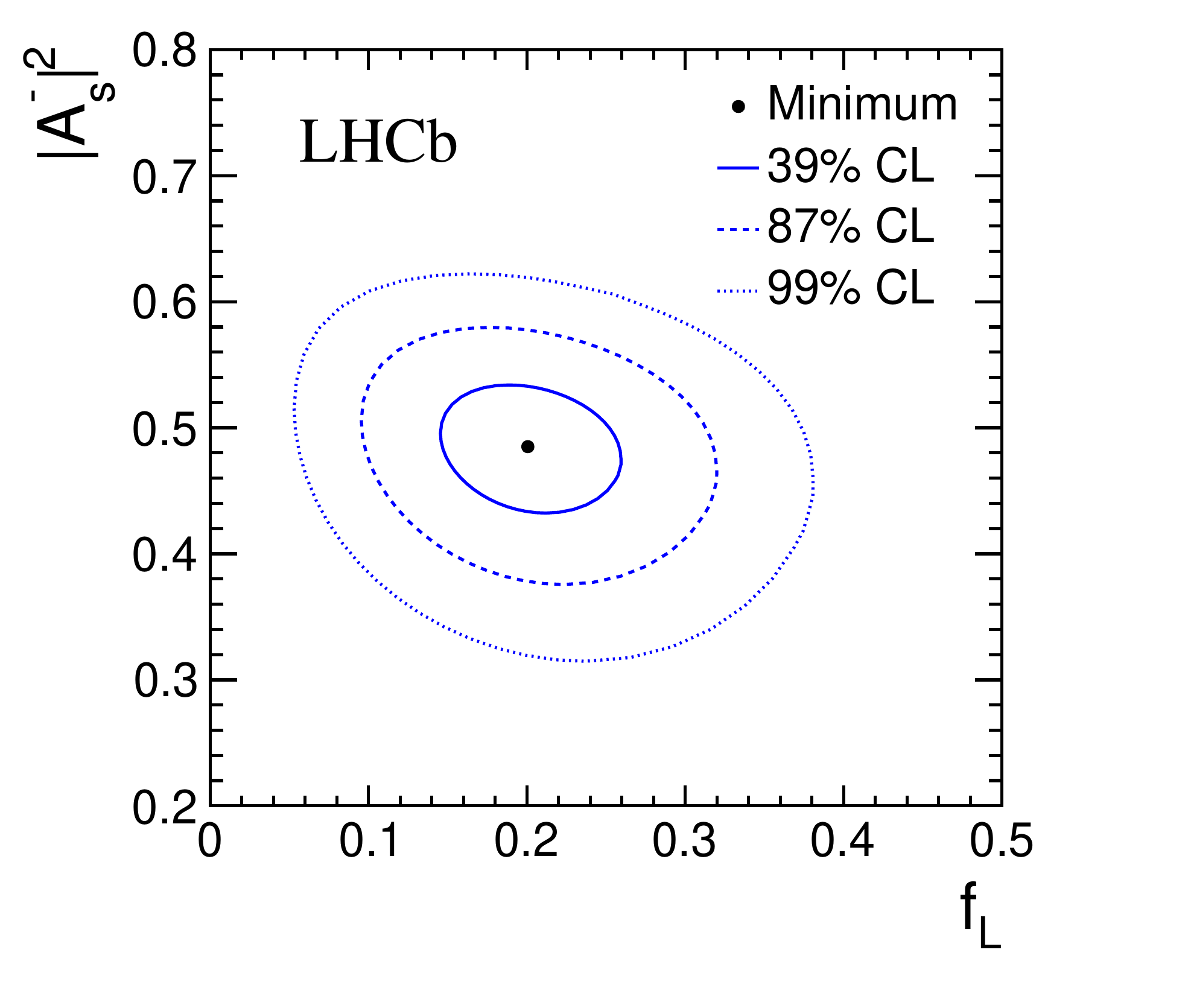}
  \end{minipage}
  \caption{(Left) Profile likelihood for the parameter $f_L$. (Right) Regions corresponding to $\Delta\ln\mathcal{L} =$ 0.5, 2 and 4.5 (39\%, 87\% and 99\% confidence level) in the $|A_s^-|^2$ -- $f_L$ plane. }
  \label{fig:contours}
\end{figure}

Figure~\ref{fig:contours} shows the likelihood for the longitudinal polarisation fraction $f_L$, where all the other 
parameters are minimised at each point of the curve, depicting parabolic behaviour around the minimum.
Additionally, confidence regions in the $|A_s^-|^2$--$f_L$ plane are shown.

\begin{table}[t]
  \caption{Results of the simultaneous fit to \BsKpiKpi TOS and non-TOS candidates with $|M(K^+, \pi^-, K^-, \pi^+) - m_{\Bs}| < 30 \mevcc$
	(phases are measured in radians). The first uncertainties are statistical and the second systematic.}
  \label{tab:fit_kstkst}
  \centering
\begin{tabular}{cc}
\hline
Parameter & Value \\
\hline
 $f_L$ &  $0.201 \pm 0.057 \pm 0.040$  \\
  $f_{\parallel}$ &  $0.215 \pm 0.046 \pm 0.015$ \\
  $|A_s^+|^2$ &  $0.114 \pm 0.037 \pm 0.023$ \\
  $|A_s^-|^2$ &  $0.485 \pm 0.051 \pm 0.019$ \\
  $|A_{ss}|^2$ &  $0.066 \pm 0.022 \pm 0.007$ \\
  $\delta_{\parallel}$ &  $5.31 \pm 0.24 \pm 0.14$ \\
  $\delta_{\perp}-\delta_s^+$ &  $1.95 \pm 0.21 \pm 0.04$ \\
  $\delta_s^-$ &  $1.79 \pm 0.19 \pm 0.19$ \\
  $\delta_{ss}$ &  $1.06 \pm 0.27 \pm 0.23$ \\
 \hline
\end{tabular}
\end{table}

\afterpage{\clearpage}

\begin{table}[p]
  \caption{Systematic uncertainties in the measurement of the magnitude and phase of the different amplitudes contributing to the \BsKpiKpi decay.}
  \centering
  \begin{tabular}{lcccccc}
    \hline
    Parameter  &  $\sigma_{\rm acc}$  &    $\sigma_{\rm sim}$  &   $\sigma_{\rm rw}$   &  $\sigma_{\rm mass}$  &  $\sigma_{\rm res}$  &   Total  \\ 
    \hline
	 $f_L$  	&  0.031	&  0.010	&  0.010	&  0.021	&  0.006   &  0.040  \\ 
	 $f_{\parallel}$  	&  0.008	&  0.008	&  0.004	&  0.005  &  0.007	   &  0.015  \\ 
	 $|A_s^+|^2$  	&  0.019	&  0.005	&  0.002	&  0.011	& 0.003   &  0.023  \\ 
	 $|A_s^-|^2$  	&  0.007	&  0.007	&  0.010	&  0.003	& 0.012   &  0.019  \\ 
	 $|A_{ss}|^2$   	&  0.003	&  0.001	&  0.000	&  0.005	& 0.003   &  0.007  \\ 
	 $\delta_{\parallel}$  	&  0.130	&  0.037	&  0.042	&  0.005	& 0.025   &  0.144  \\ 
	 $\delta_{\perp} - \delta_{s}^+$  	&  0.016	&  0.019	&  0.000	&  0.017	& 0.027   &  0.040  \\ 
	 $\delta_s^-$  	&  0.160	&  0.036	&  0.075	&  0.033	&  0.030   &  0.186  \\ 
	 $\delta_{ss}$ 	&  0.096	&  0.076	&  0.188	&  0.018	&  0.044   &  0.229   \\ 
    \hline
  \end{tabular}
\label{tab:systematics}
\end{table}


The most important systematic uncertainties in the measurement of the different amplitudes, phases and polarisation fractions 
are summarised in \tabref{tab:systematics}. They arise mainly from uncertainties in the modelling of the $K\pi$ 
mass distributions and from the assumption that the five-dimensional acceptance factorises into a product 
of two-dimensional functions. To better exploit the statistical power
of the simulated sample in the less populated regions of the phase space, e.g. the tails of the mass distribution, the angular and mass acceptances
are assumed to factorise. An alternative model is tested that allows for correlation between the angular distribution
and the $K\pi$ invariant mass, using a two-dimensional function in $(\cos\theta_i, m_i)$, universal for
\Kstarz and \Kstarzb decays. The fit is repeated with this acceptance model and a systematic uncertainty, $\sigma_{\rm acc}$ , is determined
from the variation with respect to the nominal fit result.
An additional uncertainty accounts for the limited size of the simulated samples, $\sigma_{\rm sim}$.

To test the accuracy of the simulation, kinematic distributions, such as those of the \pt of final-state particles, 
are compared between data and simulation. Since the input amplitudes used 
in the generators are different from those measured in data, an iterative method is defined
to disentangle the discrepancies associated with a different physical distribution.
This procedure supports the quality of the simulation, and allows for the determination
of the associated systematic uncertainty, $\sigma_{\rm rw}$. 

Several alternative models for the parameterisation of invariant mass propagators are used and a systematic 
uncertainty, $\sigma_{\rm mass}$, for the fit parameters is estimated from the variation of the fit results. The main contribution to this 
uncertainty comes from the \swave mass propagator, which is modelled by the LASS parameterisation~\cite{LASS} in the nominal fit.
A combination of two spin-0 relativistic Breit-Wigner distributions with the mean and width of the \kkappa and \khigh,
respectively~\cite{PDG}, and a single contribution from \khigh are also used.

Additional small uncertainties are considered to account for the effect of the invariant mass resolution, the lifetime acceptance
and possible biases induced by the fitting method ($\sigma_{\rm res}$).

\section{Measurement of {\bf\boldmath${\BRof\BsKstKst}$}}
\label{sec:br}

The branching fraction of the vector
mode \BsKstKst is updated with respect to the previous result~\cite{KstKstPaper}. 
This measurement is normalised using the \BdPhiKst decay, with $\phi\to K^+ K^-$ and $\Kstarz\to K^+ \pi^-$,
which has a topology similar to the signal decay and a well-known branching fraction.

\begin{figure}
  \centering
  \includegraphics[width=0.8\textwidth]{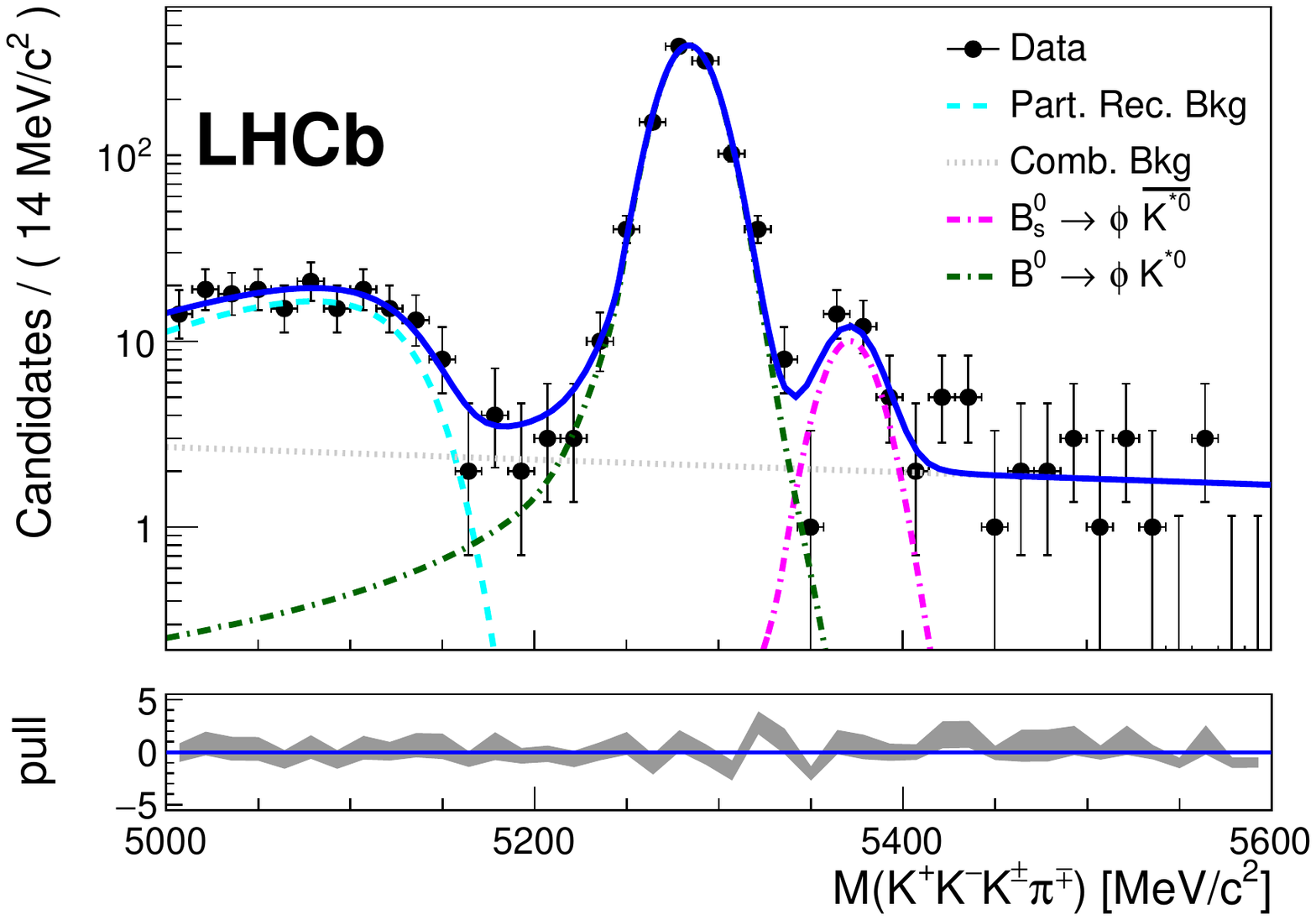}
  \caption{ Invariant mass of the selected $K^+K^-K^{\pm}\pi^{\mp}$ combinations and result of the fit to the data.
  The points represent the data and the (blue) solid line is the fit model.  The \Bs and \Bd signal peaks are shown 
  as as dashed-dotted lines (pink and dark green, respectively).  The contribution from partially
  reconstructed decays is represented as a (light blue) dashed line. The (grey) dotted line
  is the combinatorial background component. The normalised residual (pull) is shown below.}
\label{fig:refchan}
\end{figure}

The selection of \BdPhiKst decays is performed such that it closely resembles the selection of \BsKstKst decays, except for 
particle identification criteria. In particular, the requirements related to the \Bs vertex definition and the kinematic properties of 
the charged particles are identical. Figure~\ref{fig:refchan} shows the invariant mass of the final-state particles for the 
selected candidates.

The ratio of branching fractions for signal and reference decay channels is given by
\begin{eqnarray}
  \frac{\BF(\BsKstKst)}{\BF(\BdPhiKst)} &=&  \frac{\varepsilon^{\rm sel}_{\BdPhiKst}}{\varepsilon^{\rm sel}_{\BsKstKst}}  
  \times \frac{\varepsilon^{\rm trig}_{\BdPhiKst}}{\varepsilon^{\rm trig}_{\BsKstKst}}  
  \times  \frac{\lambda_{f_L}(\BdPhiKst)}{\lambda_{f_L}(\BsKstKst)}  \nonumber \\
  &&   \times  \frac{N_{\Bs} \times f_{\BsKstKst}}{ N_{\Bd} \times f_{\BdPhiKst}}
  \times   \frac{f_d}{f_s} \times \frac{\BF(\phi \rightarrow K^+ K^-)}{\BF(\Kstarz \rightarrow K^+ \pi^-)},
  \label{eq:sum_BRkstkst_ratio}
\end{eqnarray}
\noindent where $f_d/f_s$ is the ratio of probabilities for a \bquark quark to form a \Bd or a \Bs 
meson~\cite{fsfdlhcb_latest,fsfdlhcb}.

The quantities $N_{\Bs}$ and $N_{\Bd}$ represent the number of observed candidates for $\Bs \rightarrow K^+ \pi^- K^- \pi^+$
and $\Bd \rightarrow K^+ K^- K^{\pm} \pi^{\mp}$ decays, respectively, and are determined from the corresponding
fits to the four-body invariant mass spectra. The value of $N_{\Bs}$ is reported in \secref{sec:selection}.
The yield $N_{\Bd}$ is extracted from an extended unbinned maximum likelihood fit to the spectrum in \figref{fig:refchan}.
The \Bd signal is modelled by a combination of Crystal Ball and Gaussian distributions that share a common mean.
Their relative width, fraction and parameters describing the tail of the Crystal Ball function are set to the values
determined from simulation. The signal from the recently observed decay \BsPhiKst~\cite{BsPhiKst_lhcb} 
is also described using this parameterisation. 
The mass difference between \Bd and \Bs mesons is fixed to
the world average value~\cite{PDG}. The partially reconstructed background is modelled using
an ARGUS distribution with parameters free to vary in the
fit. The combinatorial background is parameterised with a decreasing exponential function. A total of $1049 \pm 33$ signal decays
for the $\Bd \rightarrow K^+ K^- K^{\pm} \pi^{\mp}$ decay are observed.

The yield of candidates corresponding to the
resonant decays, \BsKstKst and \BdPhiKst, is given by the purity factors $f_{\BsKstKst}$ and $f_{\BdPhiKst}$.
The ratio of combined reconstruction and selection efficiencies, $\varepsilon^{\rm sel}$, is
calculated using \BsKstKst and \BdPhiKst simulated events and validated using data. The inefficiency induced
by the particle identification requirements is then determined separately using large calibration samples.
The ratio of trigger efficiencies, $\varepsilon^{\rm trig}$, is computed through a data-driven method~\cite{TISTOS}.
Moreover, the overall efficiency for each channel depends on the helicity angle distribution of the final state particles,
and is encoded into the factors $\lambda_{f_L}$. Both the purity and $\lambda_{f_L}$ factors for the \BsKstKst decay are 
calculated from the results of the angular analysis. Those corresponding to \BdPhiKst decays are calculated 
from Ref.~\cite{BdPhiKst_lhcb}. 

\begin{table}[t]
  \caption{Summary of relevant quantities in the \BRof\BsKstKst calculation. The factor $\kappa(\BsKstKst)$ is
  defined as $\lambda_{f_L}(\BsKstKst)/f_{\BsKstKst}$, and equivalently for the \BdPhiKst decay.
  The first uncertainty is statistical, the second systematic.}
  \centering
  \begin{tabular}{p{0.3\textwidth}r}
    \hline
    $N_{\Bs}$ & \centering $697 \pm 31 \pm 11$\tabularnewline
    $N_{\Bd}$ & \centering $1049 \pm 33 \pm \phantom{0}7$ \tabularnewline
    $\kappa_{\BdPhiKst}/\kappa_{\BsKstKst}$ & \centering $0.453 \pm 0.059 \pm 0.040$ \tabularnewline
    $\varepsilon_{\BdPhiKst}/\varepsilon_{\BsKstKst}$ & \centering $1.30\phantom{0} \pm 0.17\phantom{0} \pm 0.07\phantom{0}$ \tabularnewline
    \hline
  \end{tabular}
\label{tab:factors_br}
\end{table}

With the factors summarised in \tabref{tab:factors_br}, the ratio of branching fractions is determined to be
\begin{equation}
\frac{\BRof\BsKstKst}{\BRof\BdPhiKst} = 1.11 \pm 0.22 ({\rm stat.}) \pm 0.12 ({\rm syst.}) \pm 0.06 (f_d/f_s).
\end{equation}
Using the average $\BRof\BdPhiKst = (9.73 \pm 0.72)\times 10^{-6}$ from the
\babar \cite{bdphikst_babar} and \belle \cite{bdphikst_belle} measurements\footnote{The measurement from \cleo~\cite{bdphikst_cleo} is excluded from this
average since \swave contributions were not subtracted in the determination of the branching fraction.}, 
corrected to take into account
different rates of $B^+B^-$ and $B^0 \bar{B}^0$ pair production from $\Upsilon(4S)$
using ${\Gamma(B^+B^-)}/{\Gamma(B^0 \bar{B}^0)} = 1.055 \pm 0.025$~\cite{PDG},
the result obtained is
\begin{equation}
  \BRof\BsKstKst = (10.8 \pm 2.1 \,{\rm (stat.)} \pm 1.4 \,{\rm (syst.)} \pm 0.6 \,(f_d/f_s) ) \times 10^{-6}. \nonumber
\end{equation}

The main systematic uncertainties considered are related to the invariant mass fit
used to determine the signal and reference event yields, the angular correction,
and the determination of the trigger efficiency. To determine the systematic uncertainty associated with the number 
of candidates, the fit is repeated using different models for the signal and background components. The 
largest variation is assigned as a 1.7\% systematic uncertainty. A 5\% uncertainty is attributed 
to the trigger efficiency, after calibration of the data-driven method applied to both channels using fully
simulated events.
The systematic uncertainty associated with the angular correction $\lambda_{f_L}$
is the result of the propagation of the systematic uncertainties evaluated for the parameters measured in the
angular analysis (9\%).
 
This result supersedes the previous measurement~\cite{KstKstPaper}, which used a less sophisticated 
estimate of the \swave contribution. If rescaled to the same \swave fraction, both results are compatible.

As a result of $\Bs$--$\Bsb$ mixing, the time-integrated flavour-averaged branching fraction ($\BR$) reported here cannot be directly compared 
with theoretical predictions formulated in terms of the decay amplitudes at $t=0$ ($\BR_0$). The relation between these branching fractions 
is given by~\cite{deBruyn}
\begin{equation}
\BR \; = \; f_{\Delta\Gamma} \; \BR_0  ,
\quad
\text{with}
\quad
f_{\Delta\Gamma} \; = \; \left( 1 - \frac{\Delta\Gamma_s}{2\Gamma_s}\left(f_L + f_{\parallel} + f_{\perp} \right) \right) \; .
\end{equation}
Using the decay widths measured in Ref.~\cite{LHCb_DeltaGamma} and the polarisation fractions reported here, the correction factor is calculated 
to be $f_{\Delta\Gamma} = 1.015\pm0.010$.

\section{Conclusions}
\label{sec:conclusions}

The decay \BsKpiKpi is studied using $pp$ collision data 
recorded by LHCb during 2011 at a
centre-of-mass energy $\sqrt{s} = 7$\tev. This sample corresponds to an integrated luminosity of 1.0\invfb.

A test of the SM is performed by measuring eight \CP-violating quantities which are predicted to be small in the SM.
All of these are found to be 
compatible with the SM expectation, within $2\sigma$ uncertainty. In addition, assuming no \CP
violation, the angular distribution
of the decay products is analysed as a function of the $K\pi$ pair invariant mass
to measure the polarisation fractions of the decay \BsKstKst as well as
the magnitude and phase of the various \swave amplitudes.
The low polarisation 
of the vector-vector decay is confirmed by the measurement
$f_L = 0.201 \pm 0.057 \,{\rm(stat.)} \pm 0.040 \,{\rm (syst.)}$, and
a large \swave contribution is found.

Finally, an update of the \BsKstKst branching fraction,
using the \BdPhiKst decay as normalisation channel, yields
$\BRof{\BsKstKst} = (10.8 \pm 2.1 \,{\rm (stat.)} \pm 1.4 \,{\rm (syst.)} \pm 0.6 \,(f_d/f_s))\times 10^{-6}$,
in agreement with the theoretical prediction~\cite{Beneke}.
This result takes into account the
\swave component measured for the first time through the angular analysis of \BsKpiKpi decays
and supersedes the measurement in Ref.~\cite{KstKstPaper}.

\section*{Acknowledgements}

\noindent We express our gratitude to our colleagues in the CERN
accelerator departments for the excellent performance of the LHC. We
thank the technical and administrative staff at the LHCb
institutes. We acknowledge support from CERN and from the national
agencies: CAPES, CNPq, FAPERJ and FINEP (Brazil); NSFC (China);
CNRS/IN2P3 (France); BMBF, DFG, HGF and MPG (Germany); INFN (Italy); 
FOM and NWO (The Netherlands); MNiSW and NCN (Poland); MEN/IFA (Romania); 
MinES and FANO (Russia); MinECo (Spain); SNSF and SER (Switzerland); 
NASU (Ukraine); STFC (United Kingdom); NSF (USA).
The Tier1 computing centres are supported by IN2P3 (France), KIT and BMBF 
(Germany), INFN (Italy), NWO and SURF (The Netherlands), PIC (Spain), GridPP 
(United Kingdom).
We are indebted to the communities behind the multiple open 
source software packages on which we depend. We are also thankful for the 
computing resources and the access to software R\&D tools provided by Yandex LLC (Russia).
Individual groups or members have received support from 
EPLANET, Marie Sk\l{}odowska-Curie Actions and ERC (European Union), 
Conseil g\'{e}n\'{e}ral de Haute-Savoie, Labex ENIGMASS and OCEVU, 
R\'{e}gion Auvergne (France), RFBR (Russia), XuntaGal and GENCAT (Spain), Royal Society and Royal
Commission for the Exhibition of 1851 (United Kingdom).


\addcontentsline{toc}{section}{References}
\setboolean{inbibliography}{true}
\bibliographystyle{LHCb}
\bibliography{main}

\newpage

\centerline{\large\bf LHCb collaboration}
\begin{flushleft}
\small
R.~Aaij$^{41}$, 
B.~Adeva$^{37}$, 
M.~Adinolfi$^{46}$, 
A.~Affolder$^{52}$, 
Z.~Ajaltouni$^{5}$, 
S.~Akar$^{6}$, 
J.~Albrecht$^{9}$, 
F.~Alessio$^{38}$, 
M.~Alexander$^{51}$, 
S.~Ali$^{41}$
% ...

{\footnotesize \it
% ...
$ ^{5}$Clermont Universit\'{e}, Universit\'{e} Blaise Pascal, CNRS/IN2P3, LPC, Clermont-Ferrand, France\\
$ ^{6}$CPPM, Aix-Marseille Universit\'{e}, CNRS/IN2P3, Marseille, France\\
% ...
$ ^{9}$Fakult\"{a}t Physik, Technische Universit\"{a}t Dortmund, Dortmund, Germany\\
% ...
$ ^{37}$Universidad de Santiago de Compostela, Santiago de Compostela, Spain\\
$ ^{38}$European Organization for Nuclear Research (CERN), Geneva, Switzerland\\
% ...
$ ^{41}$Nikhef National Institute for Subatomic Physics, Amsterdam, The Netherlands\\
% ...
$ ^{46}$H.H. Wills Physics Laboratory, University of Bristol, Bristol, United Kingdom\\
% ...
$ ^{51}$School of Physics and Astronomy, University of Glasgow, Glasgow, United Kingdom\\
$ ^{52}$Oliver Lodge Laboratory, University of Liverpool, Liverpool, United Kingdom\\
% ...
}
\end{flushleft}

\newpage

\end{document}